\documentclass{emulateapj}

\newcommand{\myemail}{urrutia@physics.ucdavis.edu}

\slugcomment{To appear in ApJ}

\shorttitle{HST Observations of Dust-reddened Quasars}
\shortauthors{Urrutia et al.}

\begin{document}

\title{Evidence for Quasar Activity Triggered by Galaxy Mergers in HST 
Observations of Dust-reddened Quasars}

\author{Tanya Urrutia\altaffilmark{1,2}, 
Mark Lacy \altaffilmark{3},
Robert H.\ Becker \altaffilmark{1,2}
}

\altaffiltext{1}{Department of Physics, University 
of California, One Shields Avenue, Davis, CA 95616; 
\myemail }

\altaffiltext{2}{IGPP, L-413, Lawrence Livermore National Laboratory, 
Livermore, CA 94550; bob@igpp.ucllnl.org}

\altaffiltext{3}{Spitzer Science Center, MS 314-6, California Institute of 
Technology, 1200 E.\ California Boulevard, Pasadena, CA 91125; 
mlacy@ipac.caltech.edu}

\begin{abstract}
We present Hubble ACS images of thirteen dust reddened Type-1 quasars 
selected from the FIRST/2MASS Red Quasar Survey. These quasars have high 
intrinsic luminosities after correction for dust obscuration ($-$23.5 
$\ge M_B \ge$ $-$26.2 from K-magnitude). The images show strong evidence of 
recent or ongoing interaction in eleven of the thirteen cases, even before 
the quasar nucleus is subtracted. None of the host galaxies are well fit by a 
simple elliptical profile. The fraction of quasars showing interaction is 
significantly higher than the 30\% seen in samples of host galaxies of 
normal, unobscured quasars. There is a weak correlation between the amount of 
dust reddening and the magnitude of interaction in the host galaxy, measured 
using the Gini coefficient and the Concentration index. Although few host 
galaxy studies of normal quasars are matched to ours in intrinsic quasar 
luminosity, no evidence has been found for a strong dependence of merger 
activity on host luminosity in samples of the host galaxies of normal 
quasars. We thus believe that the high merger fraction in our sample is 
related to their obscured nature, with a significant amount of reddening 
occurring in the host galaxy. The red quasar phenomenon seems to have an 
evolutionary explanation, with the young quasar spending the early part of 
its lifetime enshrouded in an interacting galaxy. This might be further 
indication of a link between AGN and starburst galaxies.
\end{abstract}


\section{Introduction}

The existence of a link between AGN and starbursts/Ultraluminous Infrared 
Galaxies (ULIRGs) in the local Universe has been discussed for a long time 
(e.g. \cite{heckman,sanders96,cid}). Further evidence for such a link comes 
from the tight correlation between nuclear black hole mass and bulge mass 
(e.g \cite{magorrian,tremaine}) and the black hole mass - host galaxy 
velocity dispersion (M-$\sigma$) relation in local galaxies (e.g. 
\cite{ferrarese,gebhardt}), which suggest that star-formation and accretion 
onto black holes in galaxies are intimately connected. However, HST imaging 
of the host galaxies of luminous quasars to moderate depths showed that most 
of them have the morphology of elliptical galaxies, with little evidence for 
ongoing star formation \citep{dunlop}. At low redshift only 30\% of the host 
galaxies show obvious signs of disturbance \citep{guyon}. Although quasars 
with infrared excesses are thought to host starbursts \citep{canalizo}, these 
are a small minority of the optically-selected quasar population. 

Recent surveys have shown that optically-selected quasars comprise less than 
half of the total quasar population \citep{martinez,stern}. Optical quasar 
surveys tend to miss dust-reddened quasars which have been found either with 
infrared surveys \citep{cutri,lacy}, radio surveys \citep{postman}, hard 
X-ray surveys \citep{norman} or surveys for high-ionization narrow-line 
objects \citep{zakamska}. Therefore they represent a new and largely 
uninvestigated quasar population, which may have many members that are at an 
earlier stage in their quasar activity in which dust and gas debris from the 
merger block the view of the central AGN. This fits with population synthesis 
models of the X-ray background which predict that a large fraction (close to 
80\%) of the accretion process in AGN is obscured and only the most energetic 
photons ($>$ 20 keV) can penetrate the obscuring dust barrier 
\citep{gilli,ueda,gilli2}).

\begin{figure*}
\begin{center}
\plotone{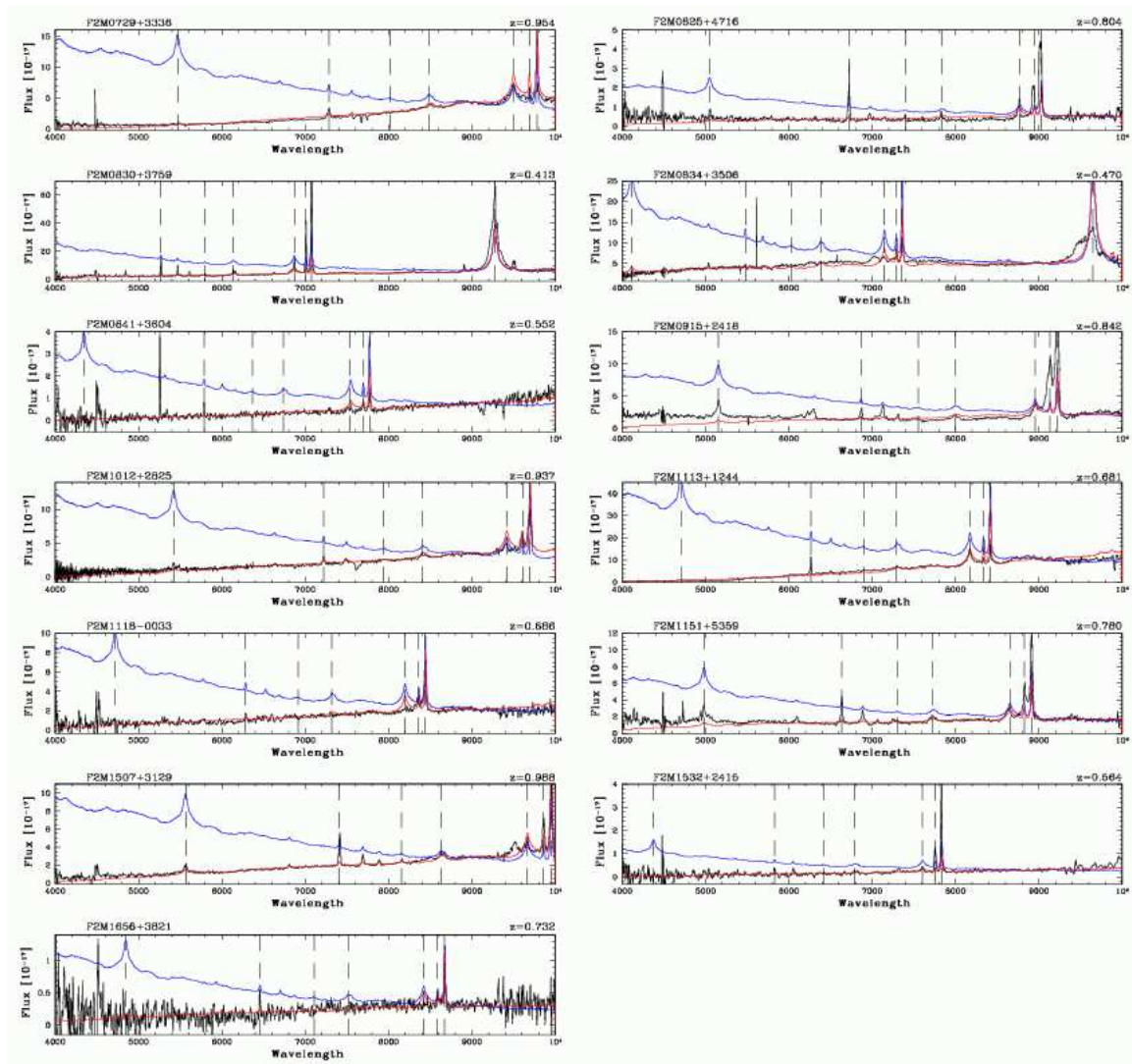}
\caption{Spectra of the FIRST/2MASS red quasars observed with HST. The black 
line in the quasar spectrum, the blue line the FBQS quasar composite 
redshifted to the quasar's redshift and the red line after applying a SMC 
reddening law to the FBQS composite.}
\label{spectra}
\end{center}
\end{figure*}

Dusty quasars are generally divided into Type-2 quasars, which show only 
narrow-line emission in the rest-frame optical, and have inferred extinctions 
towards the nucleus of $A_V \sim 5-100$, and lightly dust-reddened Type-1 
quasars, which still show broad lines and quasar continua in the optical, but 
have significant reddening relative to a normal quasars ($A_V \sim 1 - 5$).

Surveys in the mid-infrared have been able to make approximate census of the 
obscured quasar population. \cite{brown} and \cite{lacy07} find that the 
fraction of lightly dust-reddened Type-1 quasars is between 20 and 30\% of 
the total Type-1 population.  \cite{lacy07} also shows that the Type-1 
population (unreddened plus lightly reddened) is about 50\% of the total 
quasar population, the remainder being Type-2s. 

The nature of both populations remains uncertain. At present, there are two 
competing models to explain the obscuration in dusty quasars: orientation vs. 
evolution. In the orientation model, dust reddened quasars are observed 
side-on to a disk or torus of obscuring material (e.g.\cite{anton}), whereas 
in the evolution model, dusty quasars are viewed as an early stage in the 
life of a quasar, where the nucleus is obscured by debris from a merger and 
starburst which triggers the nuclear activity, prior to quasar winds 
expelling the dust (e.g. \cite{sanders}). Of particular interest are the 
class of lightly reddened quasars, where the reddening is smaller than that 
expected from a torus seen edge-on. These objects are good candidates for 
reddening by dust in the host galaxy, but may also correspond to objects in 
which the line of sight grazes the edge of the torus.

Recent simulations of galaxy mergers tend to support the evolution model, and 
suggest that the active nucleus is obscured for a long time before the 
feedback from the accretion disperses the obscuring material 
\citep{hopkins05}. In this picture, the quasar is seen in the optical regime 
only at the end of its lifetime, when it is the most luminous, radiating 
close to the Eddington limit and therefore being capable of expelling the 
dust \citep{hopkins06}. A large population of these young, obscured, 
underluminous quasars would also account for the mostly hard X-ray 
background; however there is still debate on the normalization factor of the 
hard X-ray background, that is the fraction of these sources is still 
unknown. Observations of Type-2 quasars discovered in the Sloan Digital Sky 
Survey (SDSS), however, tend to support the orientation hypothesis, with 
broad lines seen in polarized light, and host galaxies consistent with those 
of normal Type-1 quasars \citep{zakamska}. Similarly, an HST study of 
lightly-reddened Type-1 quasars selected from 2MASS found no significant 
difference in the properties of host galaxies of infrared selected quasars 
from quasars selected from other methods \citep{marble}, though ground-based 
studies of a similar samplefound more signs of interactions 
\citep{hutchings03,hutchings06}.

\begin{deluxetable*}{lccccc}
\tabletypesize\footnotesize
\tablecaption{Source properties \label{prop}}
\tablewidth{0pt}
\tablehead{ & & & \colhead{Radio Flux} & \colhead{E($B-V$)} & Luminosity \\
\colhead{Source} & \colhead{$z$} & \colhead{K-mag} & 
\colhead{1.4GHz (mJy)} & \colhead{(rest)} & \colhead{Spectrum / K-mag} }
\startdata
F2M0729$+$3336 & 0.954 & 14.5 &  3.3 & 0.83 $\pm$ 0.22 & -25.0 / -25.6 \\
F2M0825$+$4716 & 0.804 & 14.1 & 61.1 & 0.50 $\pm$ 0.35 & -21.9 / -25.8 \\
F2M0830$+$3759 & 0.413 & 14.6 &  6.4 & 0.80 $\pm$ 0.15 & -22.7 / -23.5 \\ 
F2M0834$+$3506 & 0.470 & 14.7 &  1.2 & 0.53 $\pm$ 0.10 & -22.9 / -23.7 \\
F2M0841$+$3604 & 0.552 & 14.9 &  6.5 & 1.05 $\pm$ 0.64 & -21.0 / -23.9 \\
F2M0915$+$2418 & 0.842 & 13.8 &  9.8 & 0.53 $\pm$ 0.36 & -23.7 / -26.2 \\
F2M1012$+$2825 & 0.937 & 15.2 &  9.2 & 0.66 $\pm$ 0.12 & -24.4 / -25.0 \\
F2M1113$+$1244 & 0.681 & 13.7 &  3.0 & 1.01 $\pm$ 0.24 & -25.1 / -25.5 \\
F2M1118$-$0033 & 0.686 & 14.6 &  1.3 & 0.72 $\pm$ 0.26 & -23.0 / -24.6 \\ 
F2M1151$+$5359 & 0.780 & 15.1 &  3.5 & 0.47 $\pm$ 0.13 & -23.3 / -24.7 \\
F2M1507$+$3129 & 0.988 & 15.1 &  7.8 & 0.54 $\pm$ 0.13 & -24.3 / -24.8 \\
F2M1532$+$2415 & 0.564 & 15.0 &  7.4 & 0.90 $\pm$ 0.53 & -20.5 / -23.9 \\
F2M1656$+$3821 & 0.732 & 15.1 &  4.1 & 0.55 $\pm$ 0.57 & -20.9 / -24.5  
\enddata
\end{deluxetable*}

This paper focuses on HST observations of a sample of lightly dust-reddened 
Type-1 quasars with $0.4<z<1$, selected using a combination of the FIRST 
radio survey and the 2MASS infrared survey. The primary objective is to study 
the host galaxies of red quasars and to assess if their morphologies show 
signs of recent interactions. We adopt a flat universe, $H_0$ = 70 km 
s$^{-1}$ Mpc$^{-1}$, $\Omega_{\Lambda}$ = 0.7 cosmology.

\section{Background and Sample Selection}

\subsection{The F2M sample}\label{sample}

We have selected a sample of luminous, dust-reddened quasars using a 
combination of 2MASS infrared survey \citep{2mass} and the FIRST radio survey 
\citep{first} in order to answer some of the questions posed above. The bulk 
of these dust-reddened Type-1 quasars have reddenings around $E(B-V) 
\lesssim$ 1.5, which makes them much more heavily reddened than other red 
quasar samples \citep{cutri,richards}, but they do not show the amount of 
obscuration as true Type-2 quasars. The methodology and preliminary results 
of this sample are described in \cite{f2m} and the final results in 
\cite{f2m07}. We have used two different color selection techniques, both of 
which seem to be highly effective at finding red quasars ($J - K > 1.7$ and 
$R - K > 4$; or $R - K > 5$). We have followed up our candidates with Keck 
(ESI) and IRTF (Spex) to obtain spectra in the optical and near-infrared. 
Some of these objects show only narrow lines in the observed-frame optical, 
but spectroscopy of the quasars in the near-IR have show very broad H$\alpha$ 
or Paschen lines. So far, about 100 red quasars have been found with this 
method, which would have been missed by traditional optical quasar surveys 
\citep{f2m07}. Following their conventions, objects from the FIRST-2MASS red 
quasars survey are named F2M. 

In contrast to the 2MASS quasar sample of \cite{marble}, which were mostly 
low redshift, low luminosity and selected by their near-infrared colors 
alone, the red quasars in our FIRST/2MASS study have a median redshift of 
0.7. We therefore expect the F2M sample to have have higher luminosity and 
perhaps more star-formation in the host galaxies than other red quasar 
samples at lower redshift. 

While the selection of red quasars based on $R-K$ colors could include 
objects which are red due to galaxy starlight in the infrared, the objects in 
this sample are intrinsic high luminosity quasars, in which the galaxy light 
should add a negligible contribution in the K-band. This becomes evident when 
we measure the reddening of the quasars. 

\subsection{Sample spectra}

From the F2M sample, we chose a subsample of 13 0.4 $<$ z $<$ 1.0 red quasars 
to image to inspect their host galaxy properties. In this section we will 
describe their optical spectra, taken mostly with ESI on Keck.

We fit the optical red quasar spectra to a model of an unreddened quasar. 
For that we use the FBQS composite spectrum \citep{fbqs} with a SMC dust 
reddening law from \cite{fitz99} using the relation. 

\begin{equation}
F_o(\lambda) = F_c(\lambda) \, 10^{E(B-V) \, k(\lambda)}
\end{equation}

\noindent
with $F_c(\lambda)$ the composite (FBQS) and $F_o(\lambda)$ the observed 
spectrum. Typically the SMC extinction curve $k(\lambda)$ lacks a significant 
2175 \AA\, bump seen in Galactic dust. We follow the conventions from 
\cite{fitz90} to obtain the extinction curve $k(\lambda)$, with $R_V = 3.1$ 
(Galactic value).

\begin{figure*}
\begin{center}
\includegraphics[width=5.5cm]{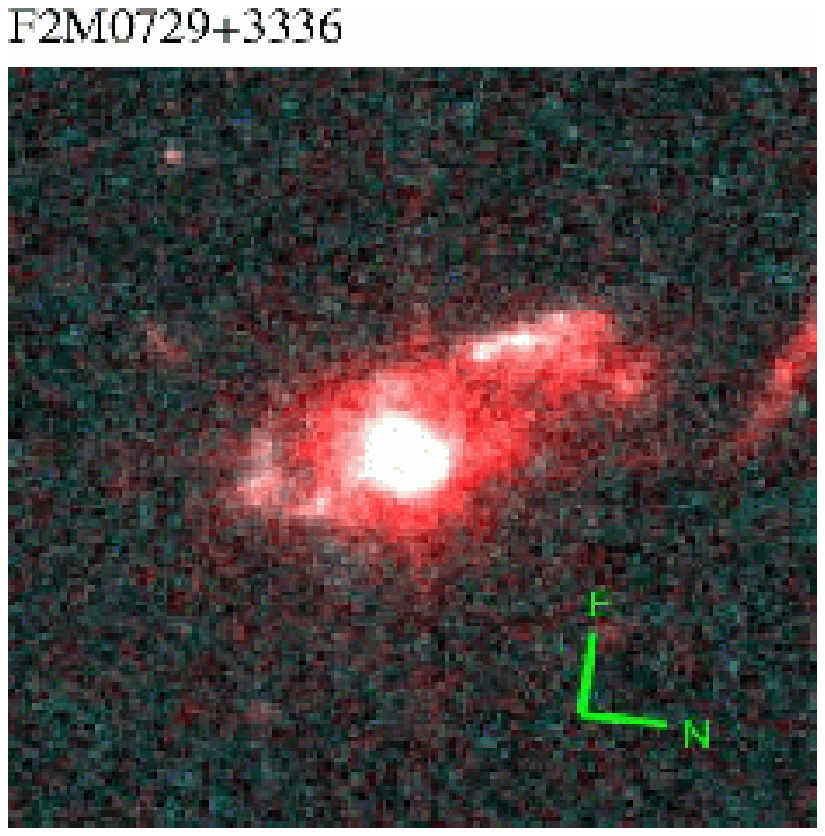}
\hspace*{1.5cm}
\vspace*{0.5cm}
\includegraphics[width=5.5cm]{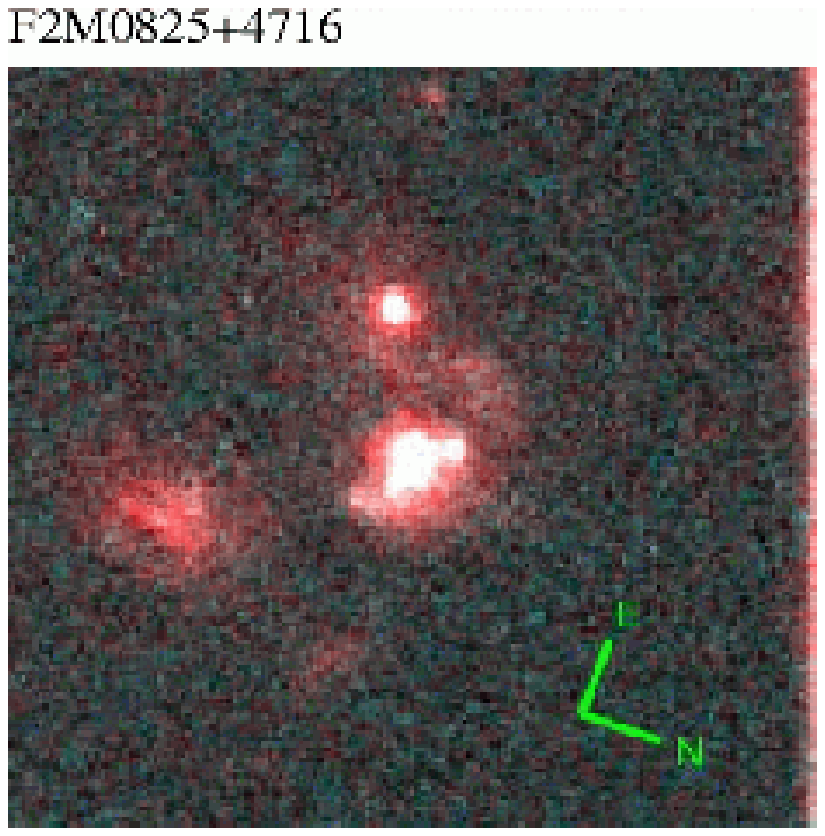}
\includegraphics[width=5.5cm]{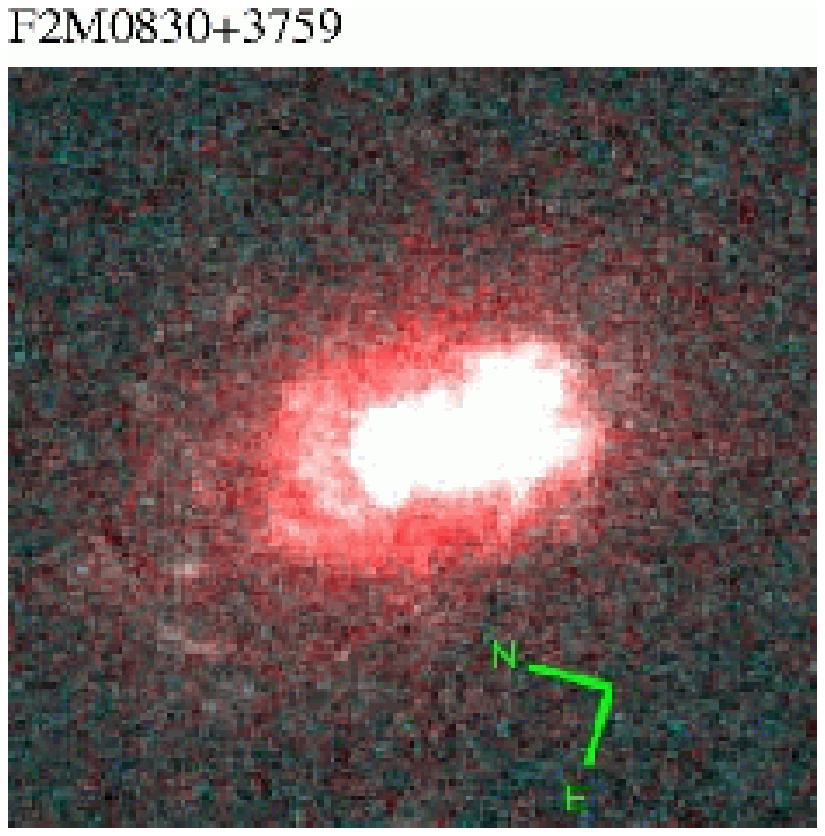}
\hspace*{1.5cm}
\vspace*{0.5cm}
\includegraphics[width=5.5cm]{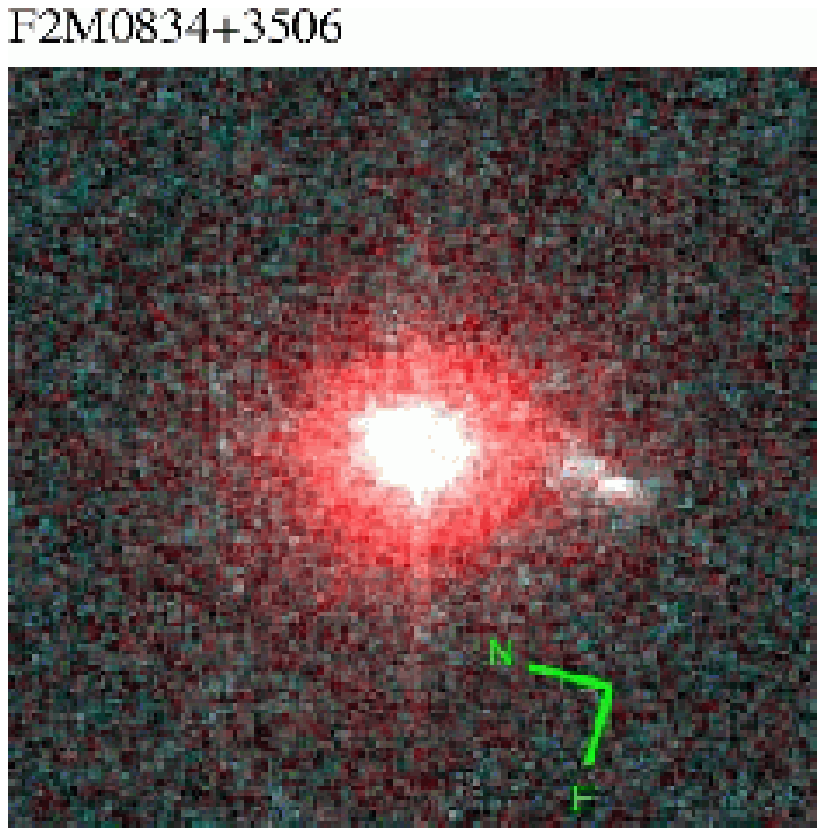}
\includegraphics[width=5.5cm]{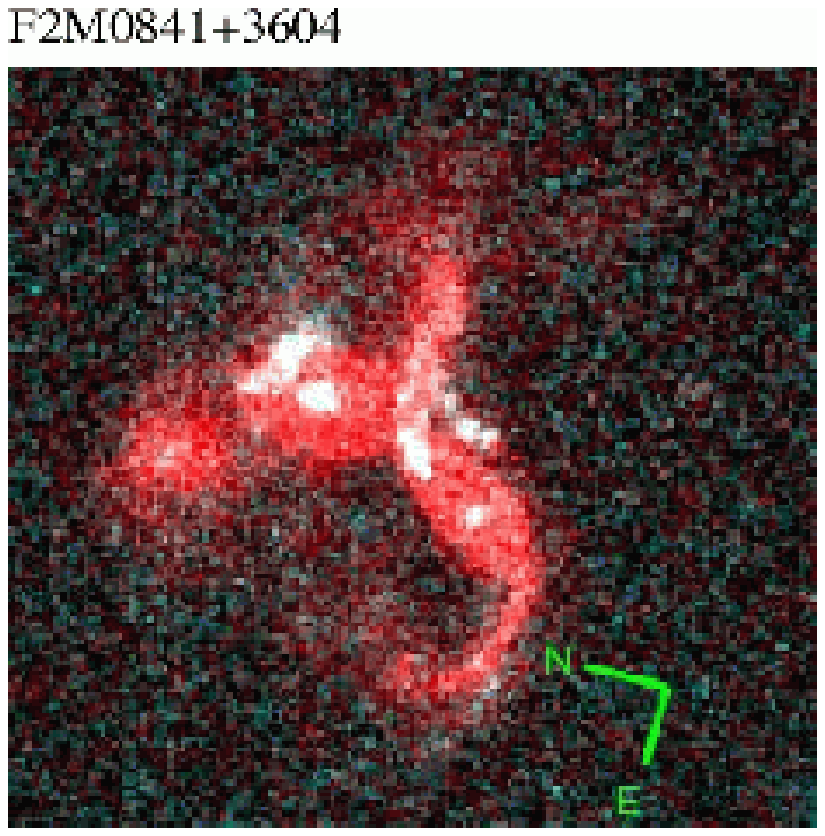}
\hspace*{1.5cm}
\includegraphics[width=5.5cm]{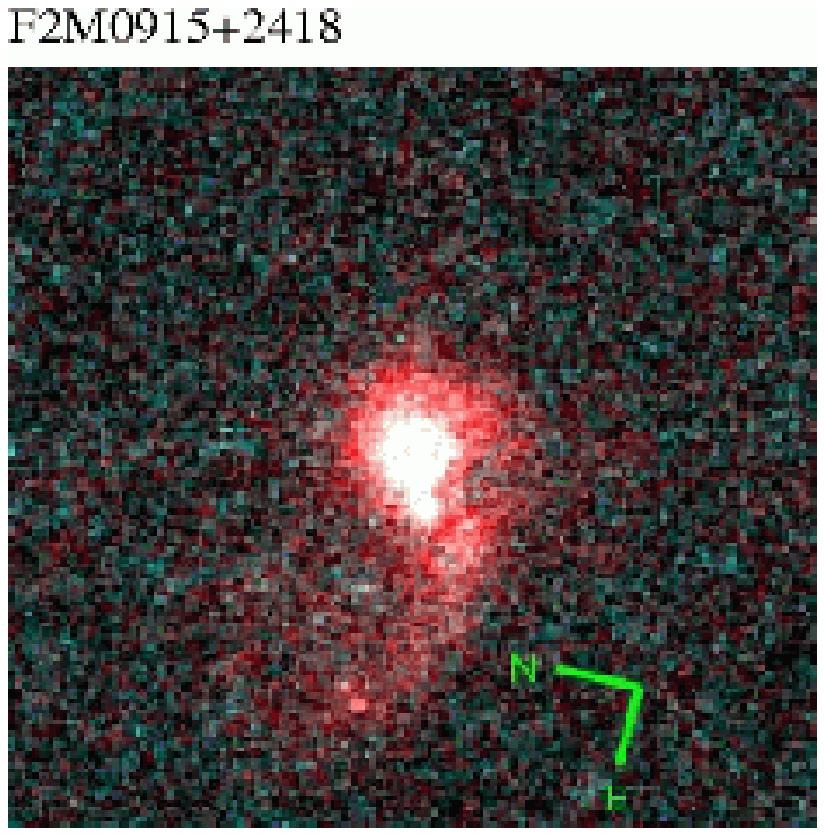}
\caption{Color composites of the 13 red quasars. We use the I-band data for 
the red part and the g-band for the blue and green parts of the composite. 
The images are $7^{''} \times 7^{''}$. The $g^{'}$-band cuts were set so that 
it the blue/green part of the image is somewhat more enhanced in comparison 
to the $I_c$-band data; the actual details from the host galaxies can be 
seen. The true color images would appear redder than shown here. For a quasar 
at redshift of z $\sim$ 0.7, which is typical for our sample, 1 arcsecond 
represents about 7 kpc.}
\label{images}
\end{center}
\end{figure*}
 
Figure \ref{spectra} shows the spectra of our quasars. The black line is the 
observed spectrum, the blue line is the FBQS composite redshifted and 
normalized to the quasar flux and the red line is the FBQS composite with the 
deduced reddening correction applied. The derived $E(B-V)$ are shown in Table 
\ref{prop}; the values have such high errors in $E(B-V)$ either because 
there was low signal to noise in the spectrum or the rms of the fit was not 
very good. Because there can be significant host galaxy contribution to the 
optical spectra, especially in the UV, our $E(B-V)$ estimates can serve as a 
lower limit. \cite{f2m07} also includes the near-infrared SPEX spectra to 
their fitting, so in some cases there is some discrepancy between their 
quoted values and ours. Their deduction of the reddening is beyond the scope 
of this paper, so we will only take the optical spectrum for an estimate of 
the reddening of the total system.

For most of the quasars the reddening fits are consistent with dust-reddening 
and very little red starlight from the host galaxy is required. However, this 
is not the case for F2M0825$+$4716, F2M0915$+$2418 and F2M1151$+$5359 in 
which the the fit for $E(B-V)$ broke down. These objects will be important 
later when we inspect the properties of their host galaxies. Note also that 
the reddenings are significant enough that the red colors of the quasars 
could not solely have come from a ``red'' spectral slope. Even though the F2M 
sample is radio selected, the radio fluxes of the objects (Table 
\ref{prop}) are also weak; the quasars are radio intermediate at best, so a 
strong synchrotron component showing up in the infrared is unlikely to have 
much contribution to the red color.

\begin{figure*}
\begin{center}
\setcounter{figure}{1}
\includegraphics[width=5.5cm]{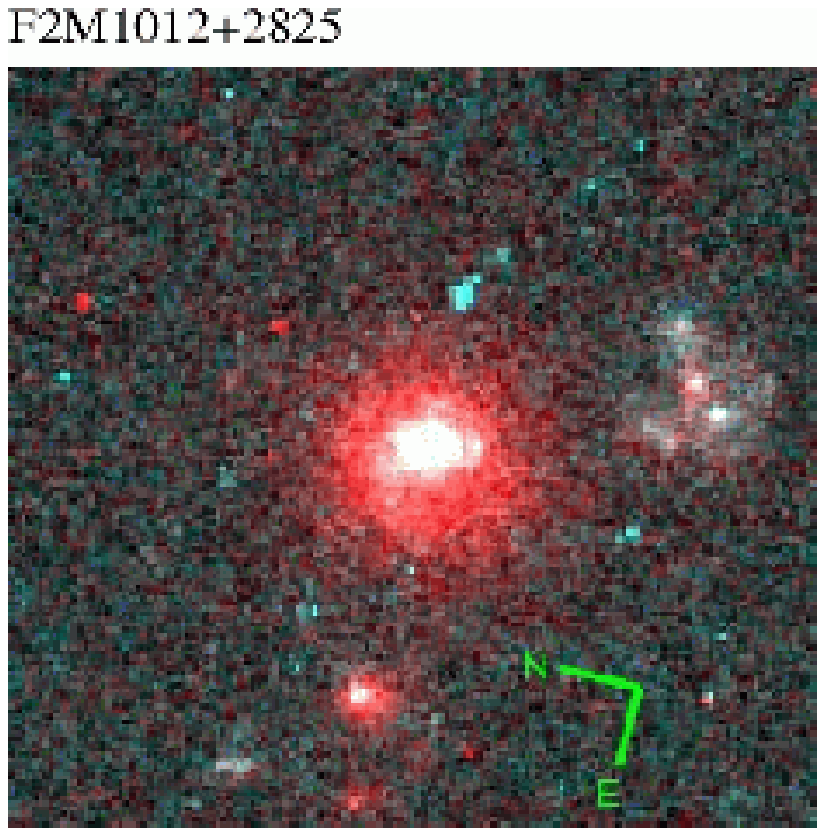}
\hspace*{1.5cm}
\vspace*{0.5cm}
\includegraphics[width=5.5cm]{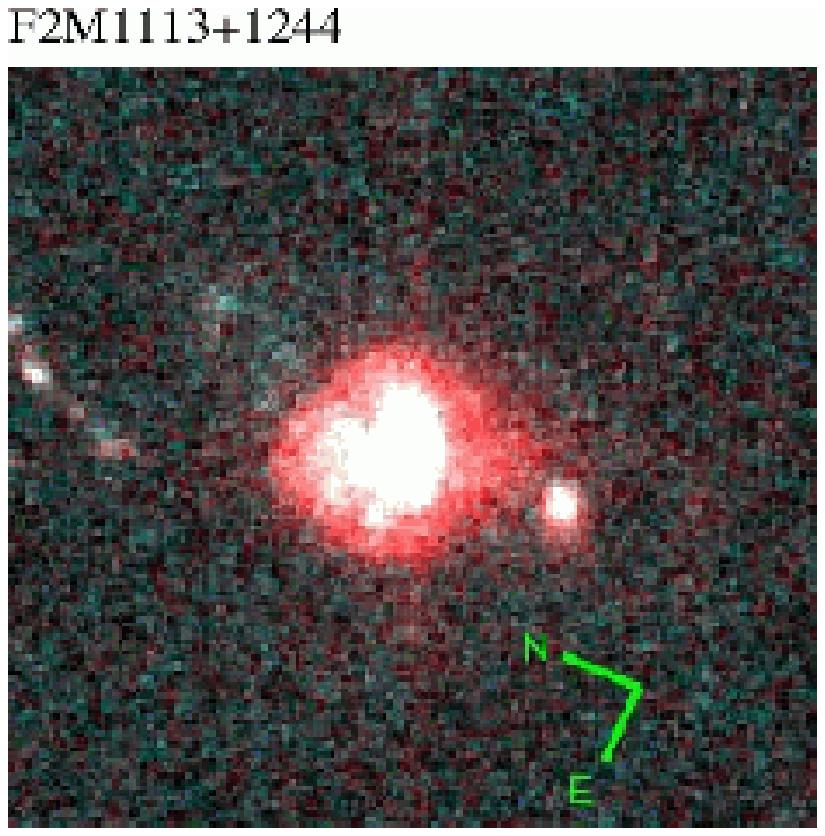}
\includegraphics[width=5.5cm]{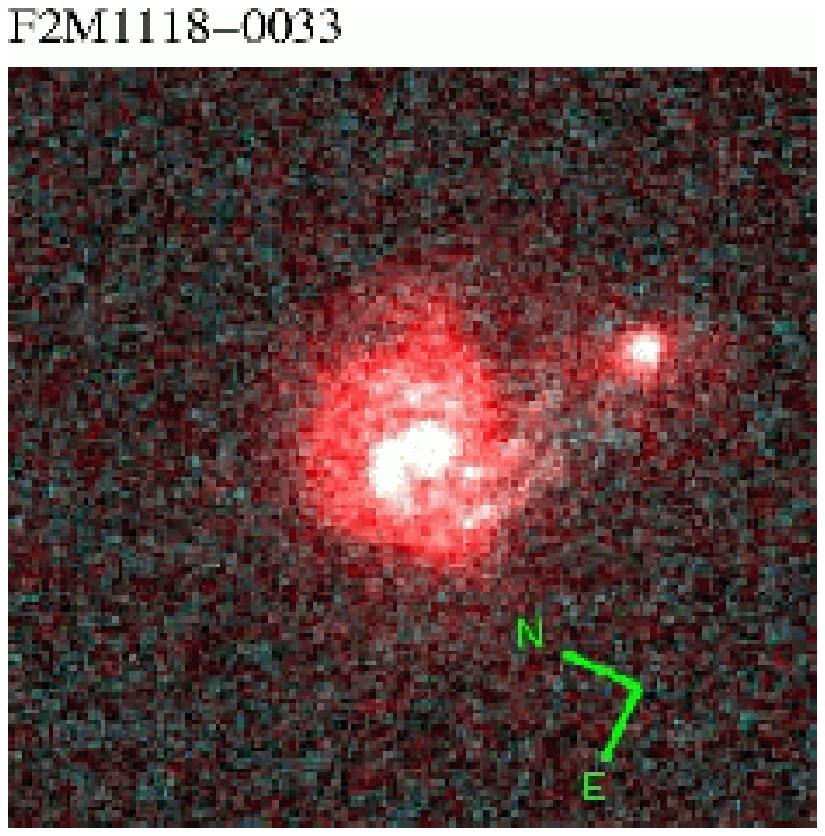}
\hspace*{1.5cm}
\vspace*{0.5cm}
\includegraphics[width=5.5cm]{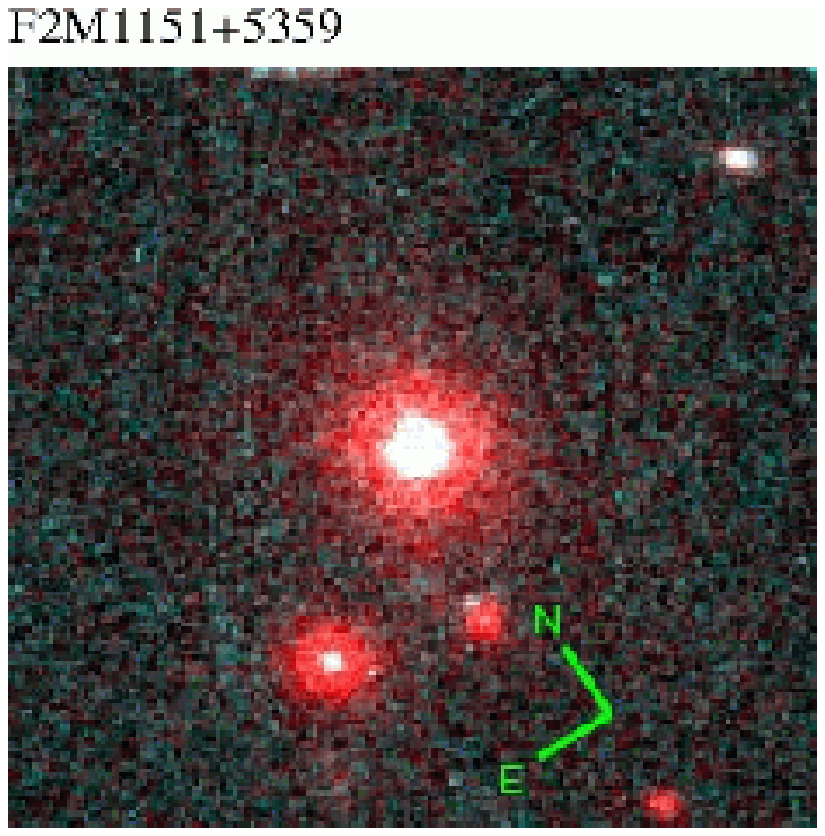}
\caption{cont.}
\end{center}
\end{figure*}

Once we have the reddening $E(B-V)$, we can also derive the intrinsic 
luminosities of the F2M quasars from the spectrum (corrected for 
obscuration). We also calculated the luminosities of the quasars from the 
K-band magnitude assuming that in the K-band there is no host galaxy 
contribution. The range between the absolute magnitude derived from the 
spectrum and from the K-band magnitude sometimes is quite large implying that 
there either were some large slit losses or that even in the K-band there is 
significant host galaxy contribution, see Table \ref{prop}. We are more 
inclined to use the derived luminosity from the K-band, since the slit will 
miss the obscured AGN often and as mentioned before, the $E(B-V)$ values are 
only a lower limit as there is host galaxy contribution to the spectrum and 
we don't have the full wavelength coverage ranging into the Mid-Infrared to 
get the true continuum shape. The fact that the K-Band magnitude is so bright 
implies that the reddenings must be larger. What we can say is that the 
quasars are more luminous than usual, having a luminosity range of $-$23.5 
$\ge M_B \ge$ $-$26.2 derived from the K-Band magnitude. Typical quasar 
luminosity functions at similar redshifts as our sample have only the highest 
luminosity quasars above $M_B \sim -24.5$ (our mean value), with a steepening 
of the function at around $M_B \sim -23.5$ for quasars at a redshift of 
$z \sim 0.7$ (e.g. \cite{qlf}). The high values of $M_B$ we find for our 
quasars suggests that we are missing the majority of this type of objects. 
Thus, because of the shallowness of the 2MASS survey we are only able to 
probe the tip of the ``red quasar iceberg''.

\section{Hubble ACS observations}\label{acs}

We followed up this subsample of 13 F2M red quasars with the ACS Wide Field 
Camera on HST (GO-10412). They were imaged with the F475W and the F814W 
filters, which roughly correspond to $g^{'}$ and $I_c$ filters. The redshift 
range was chosen such that all our objects had their intrinsic luminosities 
well above the quasar/Seyfert divide (so that the near-IR magnitudes used in 
their selection would not have much contribution from the host galaxies), and 
all were at low enough redshift that emission above rest-frame 4000 \AA\, 
fell within the bandpass of the F814W filter on ACS. The observing dates and 
corresponding exposure times are presented in Table \ref{observe}. 

\begin{figure*}
\begin{center}
\setcounter{figure}{1}
\includegraphics[width=5.5cm]{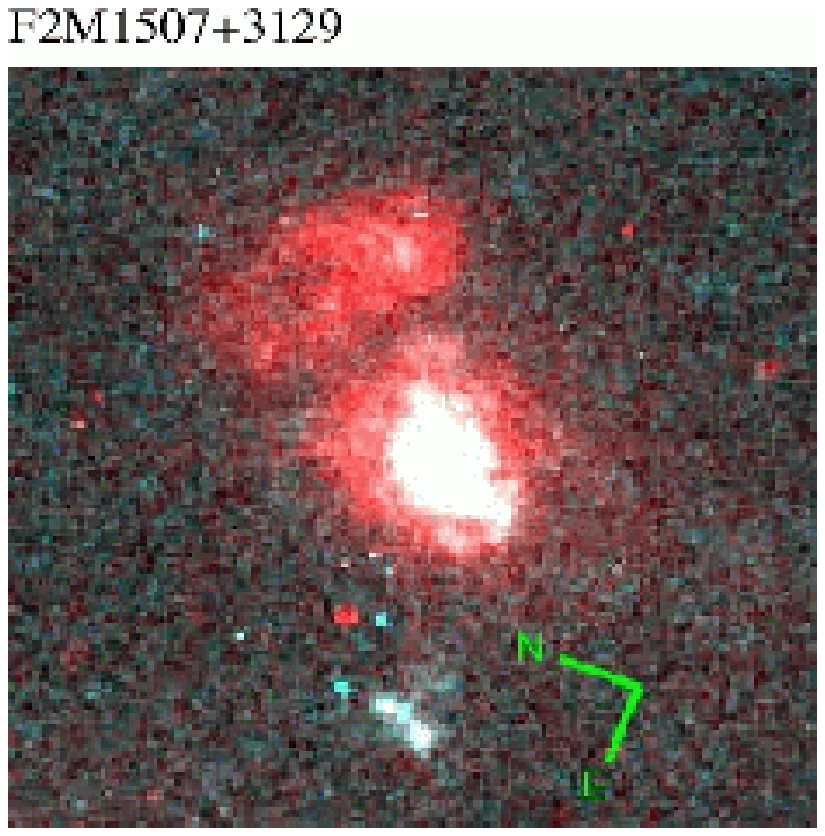}
\hspace*{1.5cm}
\vspace*{0.5cm}
\includegraphics[width=5.5cm]{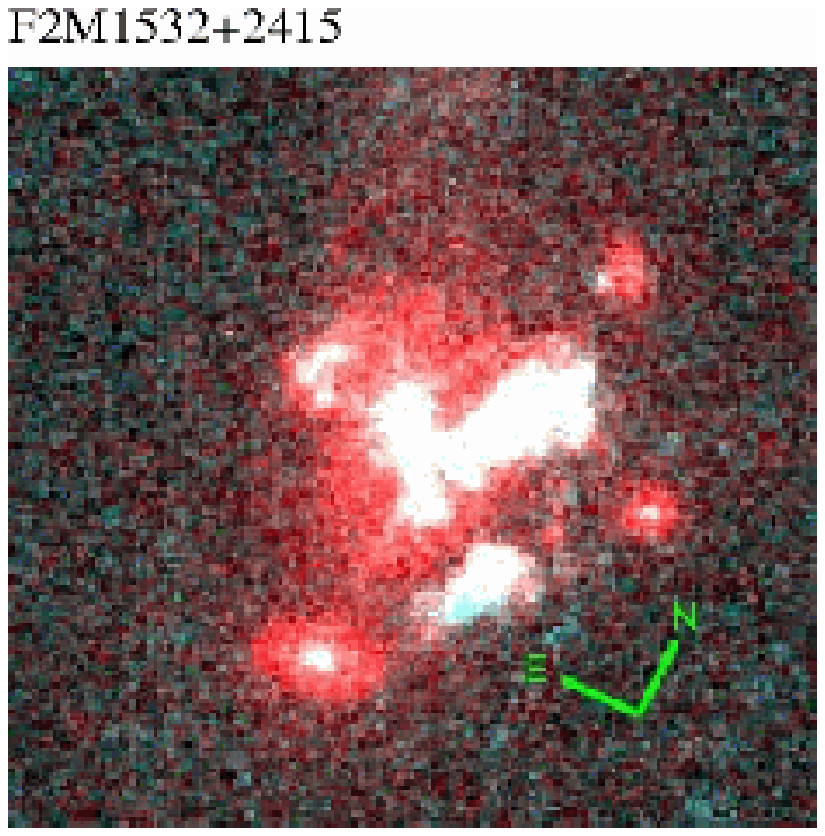}
\includegraphics[width=5.5cm]{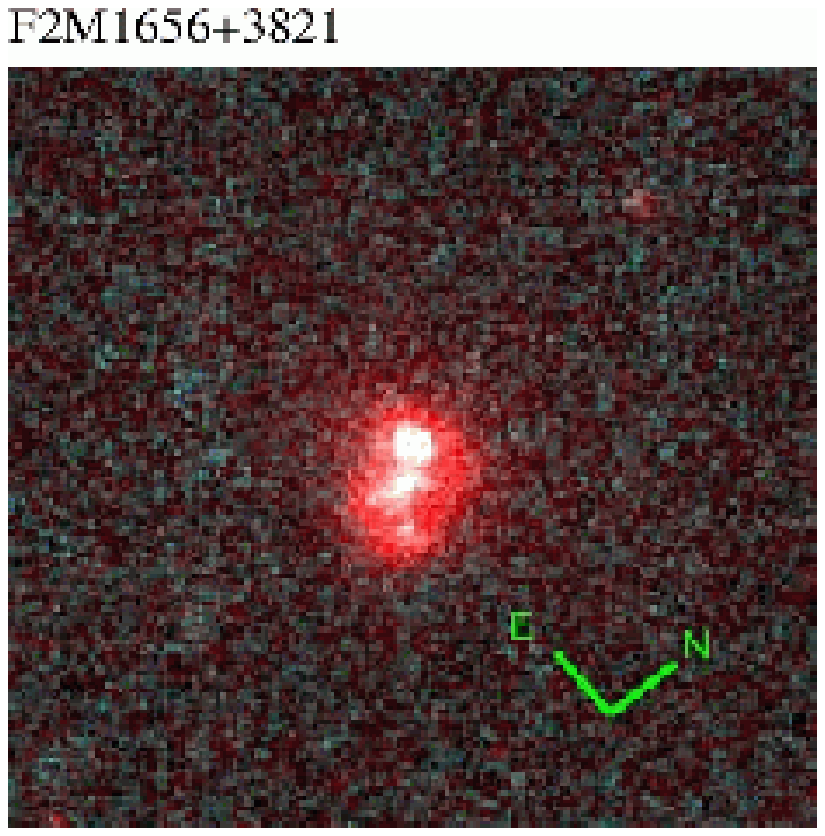}
\caption{cont.}
\end{center}
\end{figure*}

Most of the objects were observed for one HST orbit. However, as the sample 
of red quasars spanned a significant range of redshift, two high redshift 
quasars ($z > 0.9$) were observed for two orbits each to ensure a more 
uniform surface brightness limit for the sample as a whole. These images were 
drizzled with the python program {\it multidrizzle} in the usual manner. For 
the images with only one orbit we used the readily available drizzled 
products from the archive, since they did not have significant cosmic ray 
contamination.

\begin{deluxetable}{lcc}
\tabletypesize\footnotesize
\tablecaption{Journal of HST observations \label{observe}}
\tablewidth{0pt}
\tablehead{ & & \colhead{Exposure Time} \\
\colhead{Source} & \colhead{Obs. Date} & \colhead{red (s) / blue (s)} }
\startdata
F2M0729$+$3336 & Oct 24 2005 & 4650 / 4760 \\
F2M0825$+$4716 & Oct 19 2005 & 1820 / 1924  \\
F2M0830$+$3759 & Apr 18 2005 & 1720 / 1832  \\ 
F2M0834$+$3506 & Apr 18 2005 & 1620 / 1720  \\
F2M0841$+$3604 & Apr 24 2005 & 1600 / 1716  \\
F2M0915$+$2418 & Jun 05 2005 & 1672 / 1776  \\
F2M1012$+$2825 & Jun 13 2005 & 2190 / 2250  \\
F2M1113$+$1244 & May 26 2005 & 1660 / 1760  \\
F2M1118$-$0033 & Jul 14 2005 & 1640 / 1752  \\ 
F2M1151$+$5359 & Apr 20 2005 & 1860 / 1972  \\
F2M1507$+$3129 & Jul 28 2005 & 2290 / 2340  \\
F2M1532$+$2415 & Apr 24 2005 & 1736 / 1840  \\
F2M1656$+$3821 & Apr 19 2005 & 1760 / 1876 
\enddata
\end{deluxetable}

Color composite images of the red quasars are shown in Figure \ref{images}. 
Inspecting the images by eye, one can see much more information of the host 
galaxy than one would from a blue quasar in which the quasar light dominates. 
However, the images reveal much more interaction than the usual fraction of 
30\% \citep{marble,guyon} with tidal tails and many irregularities present in 
the host galaxies.

Some of the images show several compact knots or nuclei. If the merger was 
driven by two galaxies with central black holes, it is possible that the two 
of them ignite as AGN, before merging into a massive active galaxy with only 
one black hole in the center. This has been observed in NGC6240, an 
Ultraluminous Infrared Galaxy at very low redshift \citep{ngc6240}. Table 
\ref{double} summarizes the properties of the quasars that show more than one 
candidate nucleus. F2M0841$+$3604, which has the largest angular separation, 
was observed with Chandra, but only had 7 (very hard) counts, which is not 
enough to resolve both components. Two of the objects also have double-peaked 
broad lines, F2M0825$+$4716 and F2M1507$+$3129. Such emission lines are not 
uncommon in broad-line radio galaxies, and are thought to be due to a 
disk-like BLR seen edge-on (e.g. \cite{era}), however, they could also be 
from two separate active nuclei close to merger. Unfortunately, neither of 
these show multiple nuclei in their images.

\begin{deluxetable*}{lcccccc}
\tabletypesize\scriptsize
\tablecaption{Comparison of total magnitudes \label{totalmags}}
\tablewidth{0pt}
\tablehead{ & \colhead{mag$_{I_c}$} & \colhead{mag$_{g^{'}}$} & 
\colhead{mag$_{I_c}$} & \colhead{mag$_{g^{'}}$} & \colhead{mag$_i$} & 
\colhead{mag$_g$} \\
\colhead{Source} & \colhead{Aperture (3'')} & \colhead{Aperture (3'')} & 
\colhead{Isophot} & \colhead{Isophot} & \colhead{SDSS (petro)} & 
\colhead{SDSS (petro)} }
\startdata
F2M0729$+$3336 & 19.17 $\pm$ 0.05 & 21.71 $\pm$ 0.15 & 19.20 $\pm$ 0.05 & 
21.82 $\pm$ 0.15 & -    & - \\
F2M0825$+$4716 & 21.04 $\pm$ 0.11 & 22.45 $\pm$ 0.21 & 21.05 $\pm$ 0.11 & 
22.50 $\pm$ 0.21 & 18.93 $\pm$ 0.08 & 21.12 $\pm$ 0.17 \\
F2M0830$+$3759 & 18.45 $\pm$ 0.03 & 20.24 $\pm$ 0.07 & 18.38 $\pm$ 0.03 & 
20.19 $\pm$ 0.07 & 18.33 $\pm$ 0.03 & 20.06 $\pm$ 0.04 \\ 
F2M0834$+$3506 & 18.57 $\pm$ 0.04 & 20.31 $\pm$ 0.08 & 18.57 $\pm$ 0.04 & 
20.39 $\pm$ 0.08 & 19.07 $\pm$ 0.03 & 20.98 $\pm$ 0.08 \\
F2M0841$+$3604$^a$ & 20.45 $\pm$ 0.09 & 24.06 $\pm$ 0.44 & 20.23 $\pm$ 0.08 & 
25.53 $\pm$ 0.51 & 20.10 $\pm$ 0.09 & 21.98 $\pm$ 0.23 \\
F2M0915$+$2418 & 19.53 $\pm$ 0.06 & 20.68 $\pm$ 0.09 & 19.56 $\pm$ 0.06 & 
20.77 $\pm$ 0.09 & 19.95 $\pm$ 0.06 & 20.63 $\pm$ 0.04 \\
F2M1012$+$2825 & 20.23 $\pm$ 0.08 & 22.67 $\pm$ 0.23 & 20.24 $\pm$ 0.08 & 
22.77 $\pm$ 0.24 & 20.38 $\pm$ 0.63 & 22.66 $\pm$ 0.64 \\
F2M1113$+$1244 & 18.60 $\pm$ 0.04 & 21.02 $\pm$ 0.11 & 18.64 $\pm$ 0.04 & 
21.03 $\pm$ 0.11 & 18.78 $\pm$ 0.03 & 20.78 $\pm$ 0.07 \\
F2M1118$-$0033 & 19.46 $\pm$ 0.06 & 21.71 $\pm$ 0.15 & 19.31 $\pm$ 0.05 & 
21.69 $\pm$ 0.15 & 19.41 $\pm$ 0.06 & 21.49 $\pm$ 0.13 \\ 
F2M1151$+$5359 & 20.05 $\pm$ 0.07 & 21.41 $\pm$ 0.13 & 20.05 $\pm$ 0.07 & 
21.50 $\pm$ 0.13 & 20.06 $\pm$ 0.11 & 21.31 $\pm$ 0.10 \\
F2M1507$+$3129 & 19.48 $\pm$ 0.06 & 21.73 $\pm$ 0.15 & 19.39 $\pm$ 0.05 & 
21.80 $\pm$ 0.15 & 19.73 $\pm$ 0.12 & 21.66 $\pm$ 0.13 \\
F2M1532$+$2415$^b$ & 19.46 $\pm$ 0.06 & 21.94 $\pm$ 0.16 & 19.20 $\pm$ 0.05 & 
22.06 $\pm$ 0.17 & 19.09 $\pm$ 0.06 & 20.78 $\pm$ 0.10 \\
F2M1656$+$3821 & 20.72 $\pm$ 0.10 & 23.65 $\pm$ 0.36 & 20.69 $\pm$ 0.10 & 
23.68 $\pm$ 0.36 & 20.76 $\pm$ 0.20 & 23.54 $\pm$ 0.92 \\
\enddata
\tablecomments{$^a$ HST magnitude quoted for brightest component. The 
different components are not resolved in SDSS.\\
$^b$ HST $g^{'}$-band magnitute quoted for brightest component. The different 
components are not resolved in SDSS or HST I-band.}
\end{deluxetable*}

\begin{deluxetable}{cccc}
\tablecaption{Sources with multiple components \label{double}}
\tablewidth{0pt}
\tablehead{\colhead{Source} & \colhead{Comp.} & 
\colhead{Separation} & \colhead{Comments}}
\startdata
0841$+$3604 & 2 & 1.70''/9.2 kpc & Wide separation \\
1012$+$2825 & 2 & 0.15''/1.2 kpc & Resolved in $g^{'}$-band \\
1118$-$0033 & 2 & 0.40''/2.8 kpc & One component stellar \\ 
1532$+$2415 & 2 & 0.75''/4.9 kpc & One component stellar\\
1656$+$3821 & 3 & 0.80''/5.8 kpc & 2 components stellar\\
\enddata
\end{deluxetable}

We then performed photometry measurements on the images in the two bands. For 
this we used Sextractor \citep{sextractor} with parameters set as described 
in \cite{phot}. We used the AB magnitude zeropoints from the ACS webpage. We 
extracted two magnitudes this way: first the isophotal corrected magnitude 
(assuming a symmetric Gaussian profile for the object) and then a 3'' 
aperture magnitude. A large difference between the two may indicate that a 
lot of the light comes from a galaxy component outside the nucleus. We also 
compared the magnitudes with the SDSS DR5 magnitudes. All quasars, except 
F2M0729$+$3336 had SDSS imaging. Whenever the HST and SDSS magnitudes differ 
greatly, it means that the SDSS wasn't able to separate different components 
from the actual quasar due to its poor resolution. Table \ref{totalmags} 
presents the magnitudes.

The astrometry of ACS often has errors of up to 1.5 arcseconds, so we shifted 
the WCS of the images using SDSS stars in the field as references. This 
greatly improved our astrometry, to within a radial RMS error of 0.1 
arcseconds. We then compared these newly calibrated images with the FIRST 
data to confirm that the radio emission comes from the central nucleus or, in 
some cases, locate the likely position of the nucleus. The FIRST and SDSS 
surveys are fit to the same astrometric system and both have position errors 
less than 0.2 arcseconds. In most cases the radio emission is associated with 
the brightest central optical component. For F2M1656$+$3821, the radio 
emission comes from the brightest of the three components. Yet for 
F2M0841$+$3604 the radio emission emanates from the center between the two 
brightest points. The lower $\sigma$ contours are skewed a little towards the 
brightest source. It is not clear where the radio emission is coming from in 
this source.

\section{Analysis and Results}

\subsection{Properties of the Quasars}\label{propqso}

\begin{figure*}
\begin{center}
\vspace*{-0.5cm}
\includegraphics[width=12.3cm]{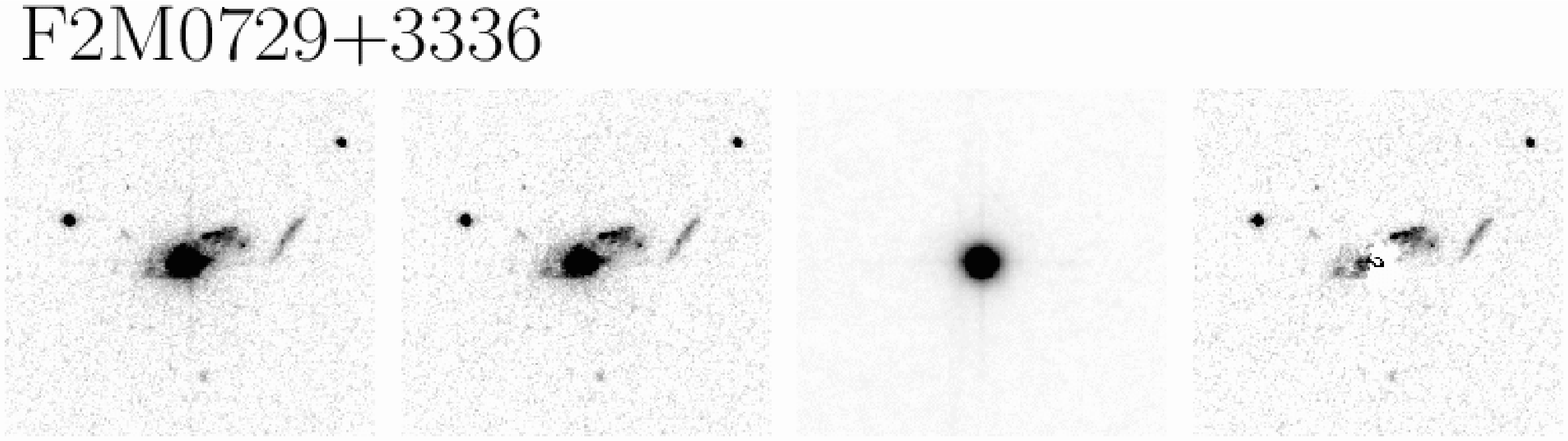}
\vspace*{0.5cm}
\includegraphics[width=4cm]{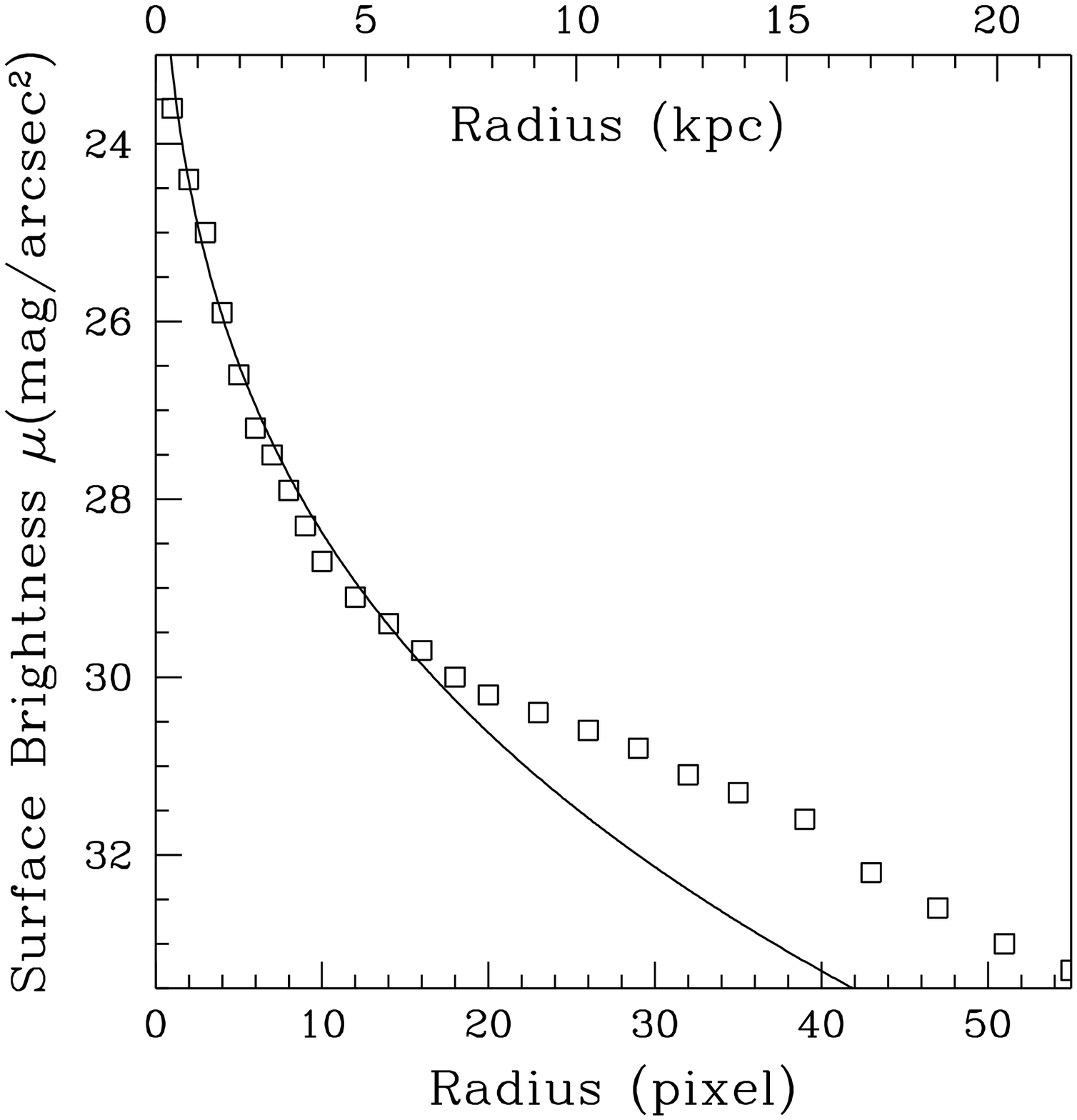}
\includegraphics[width=12.3cm]{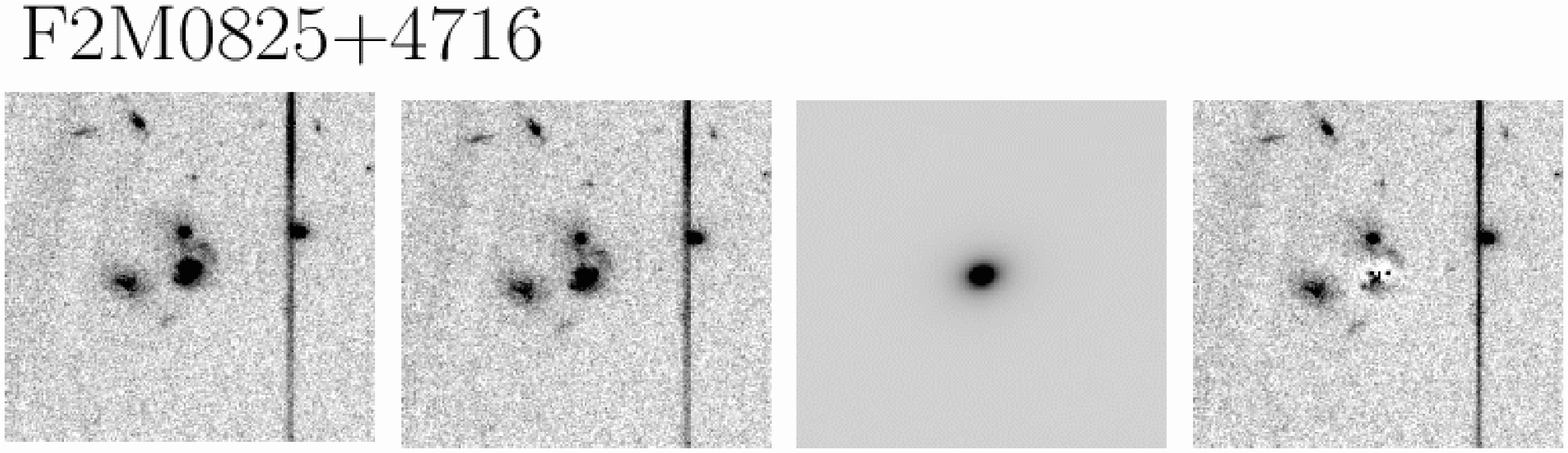}
\vspace*{0.5cm}
\includegraphics[width=4cm]{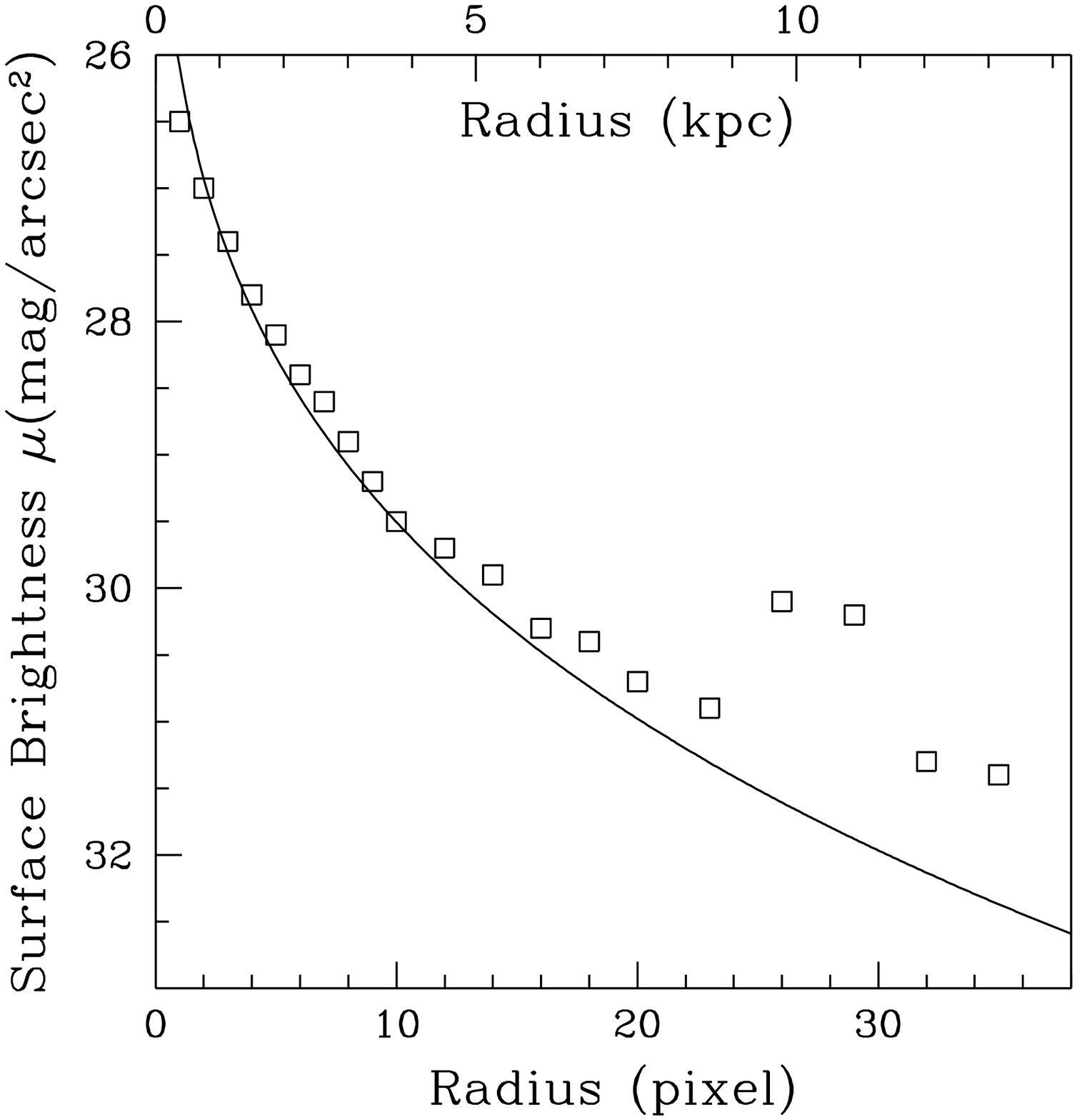}
\includegraphics[width=12.3cm]{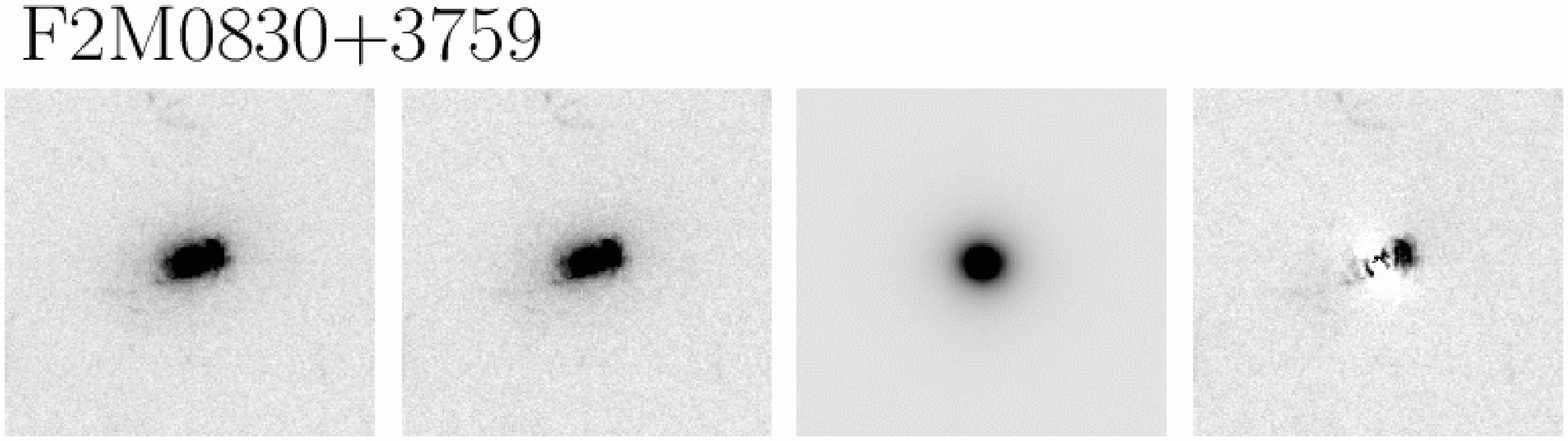}
\vspace*{0.5cm}
\includegraphics[width=4cm]{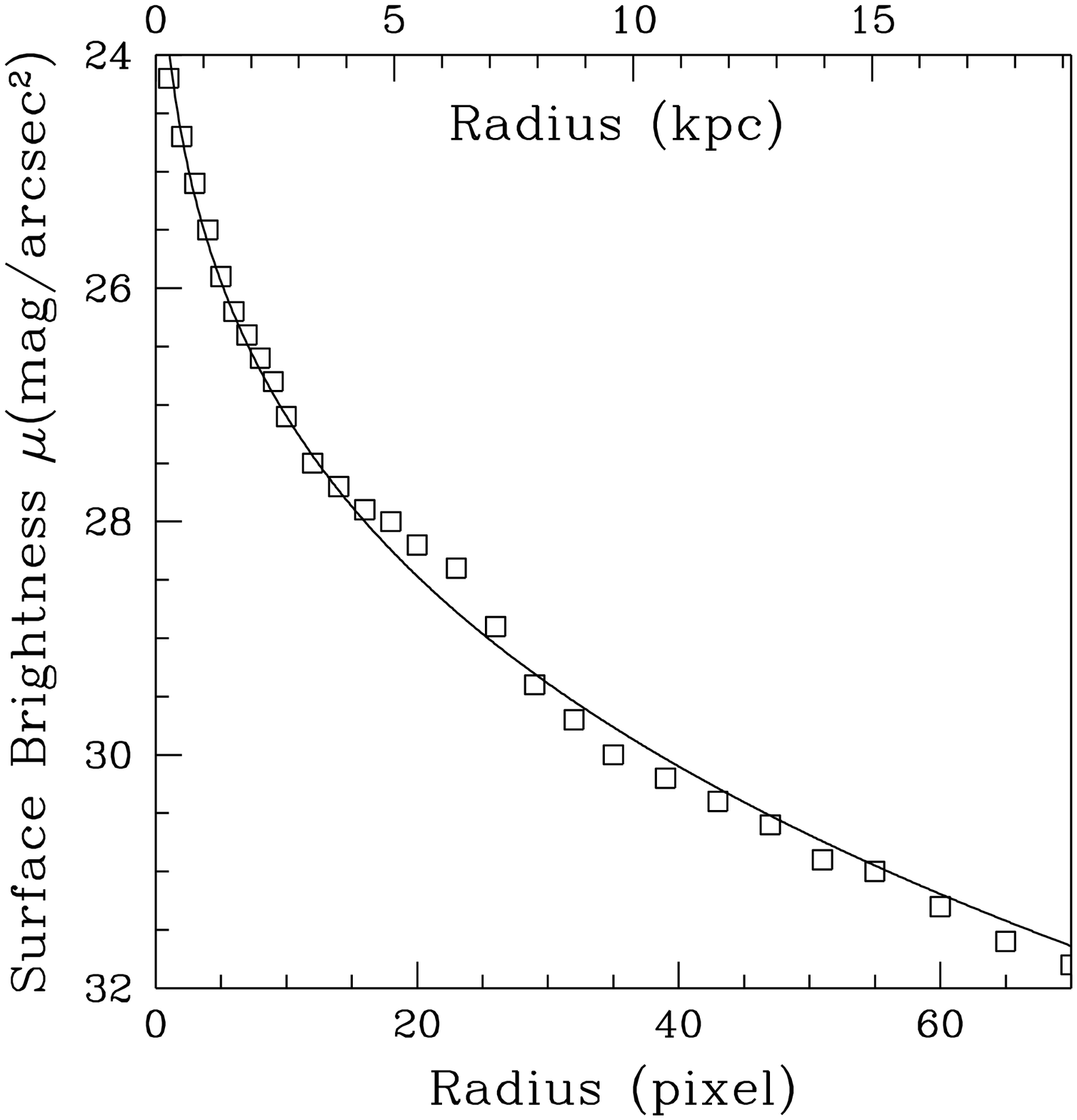}
\includegraphics[width=12.3cm]{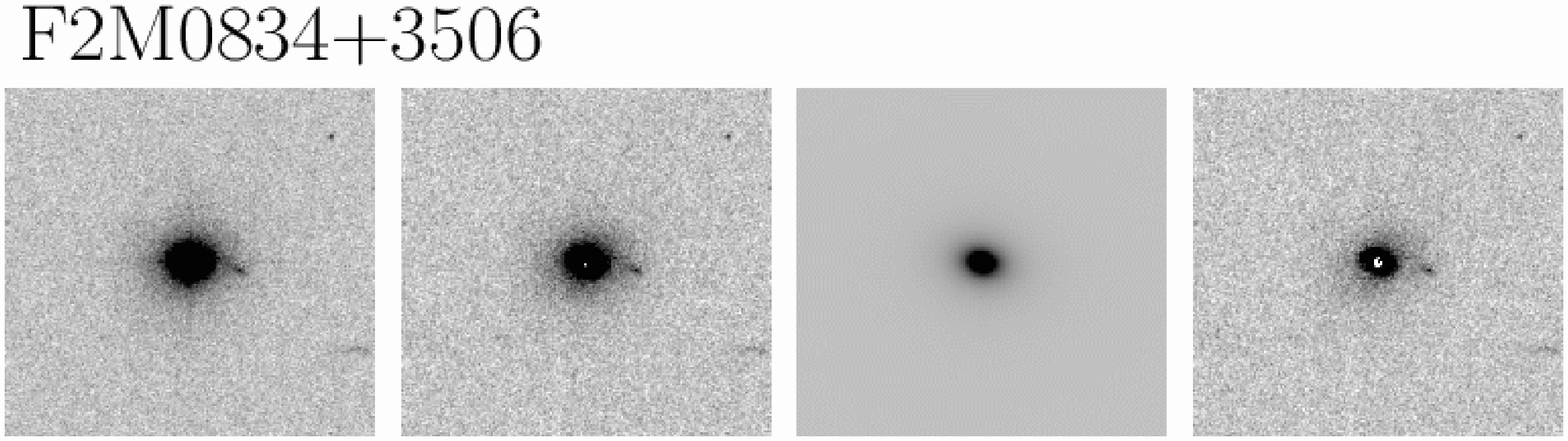}
\vspace*{0.5cm}
\includegraphics[width=4cm]{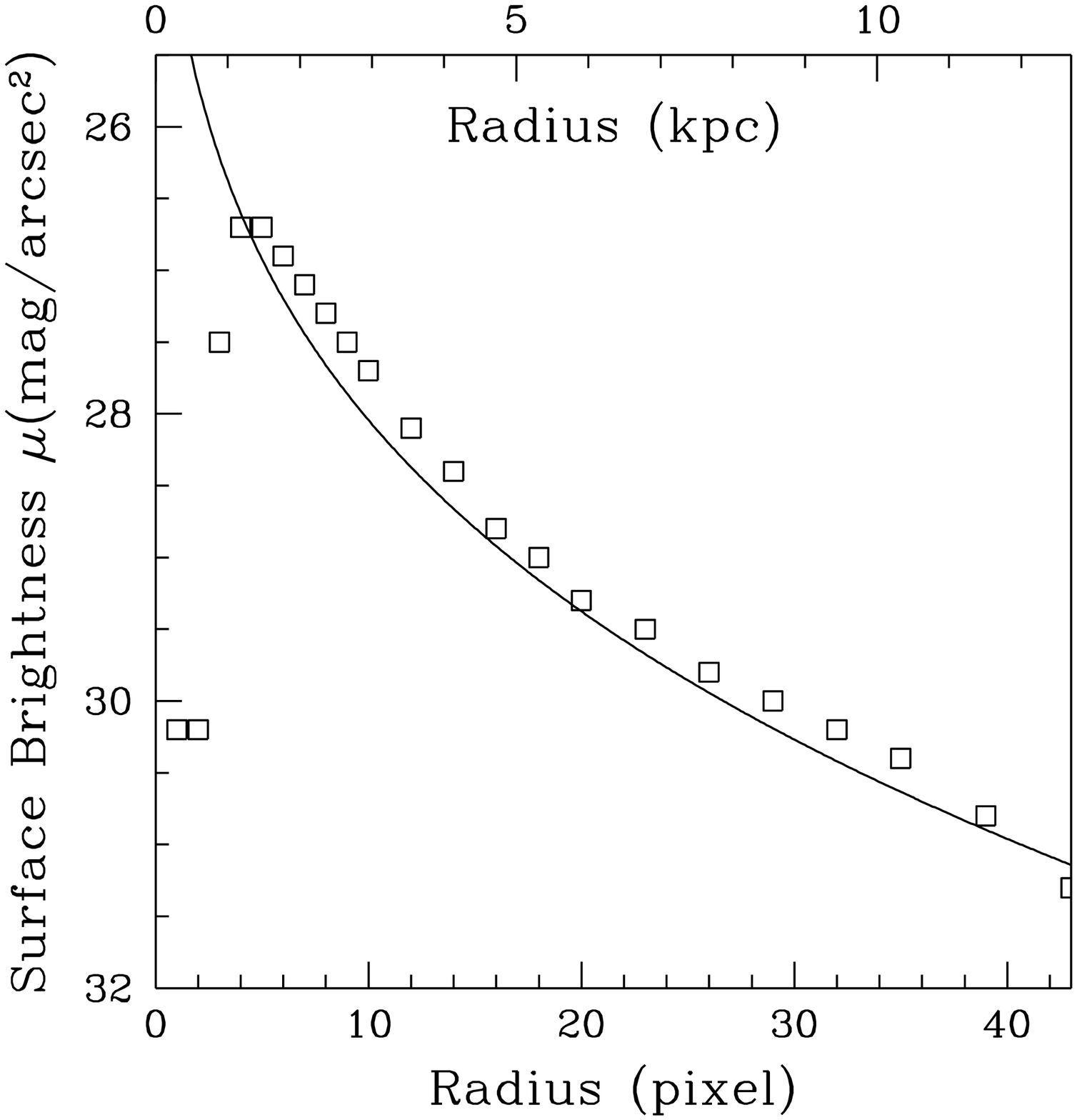}
\includegraphics[width=12.3cm]{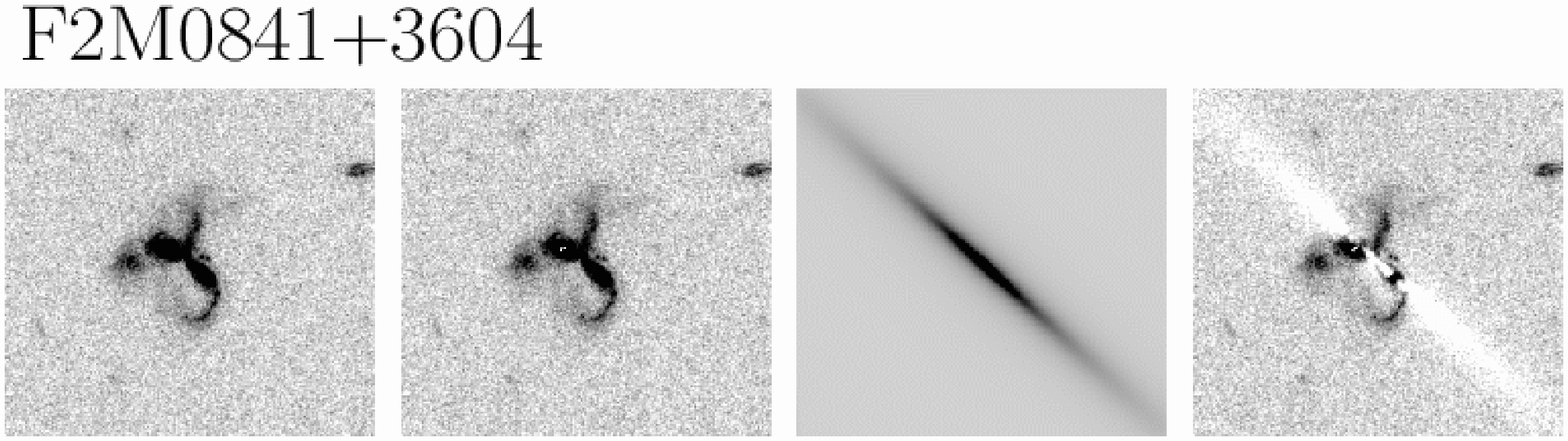}
\includegraphics[width=4cm]{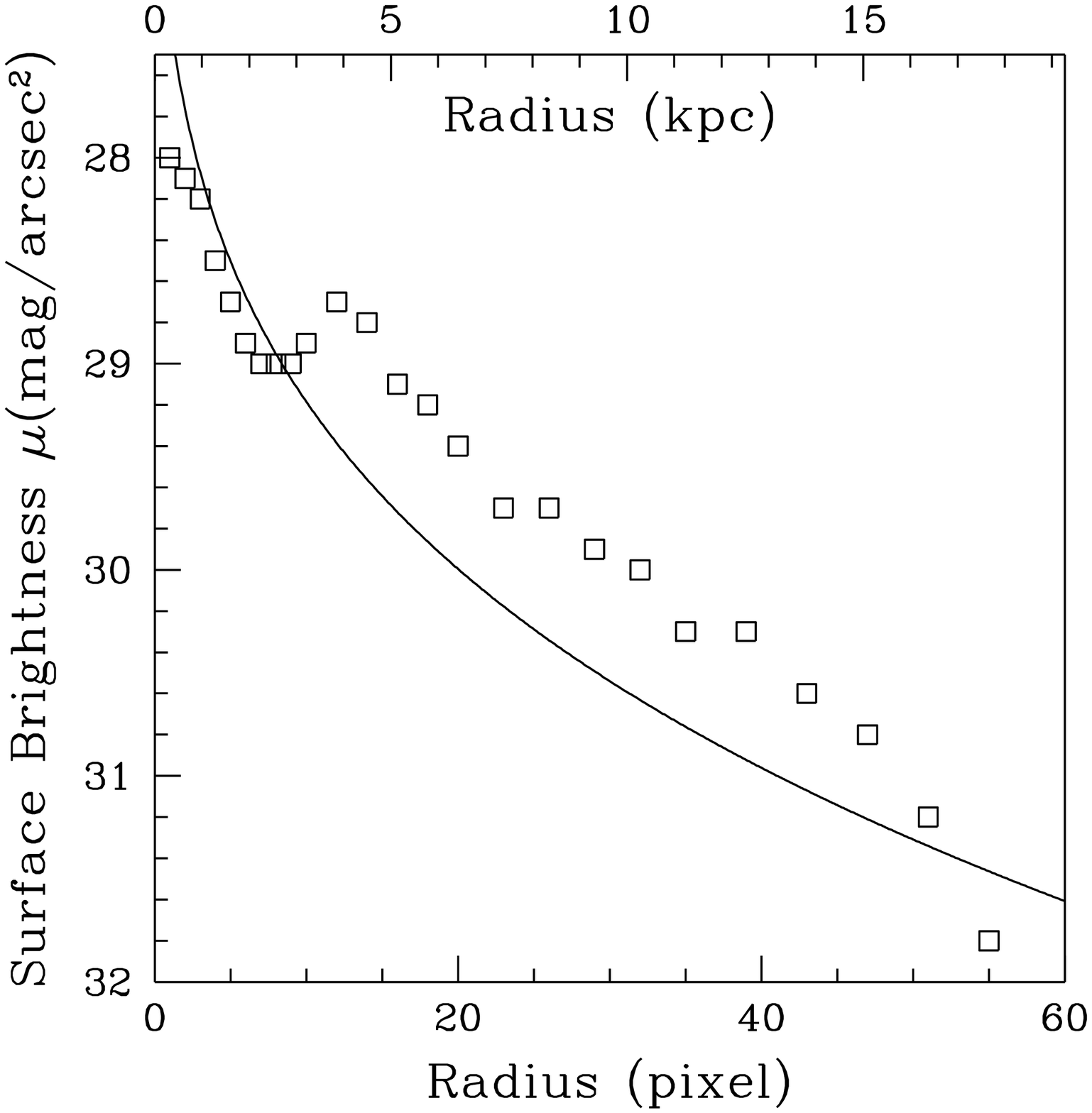}
\caption{$I_c$-band results of PSF and Host Galaxy fitting. The first column 
shows an original postage stamp of the red quasar image, the second column 
the PSF-subtracted image, the third the best-fit elliptical model, the fourth 
the residual after subtracting the model and the fifth the radial surface 
brightness profile with the solid lines representing the elliptical fit.}
\label{subtraction}
\end{center}
\end{figure*}

We performed quasar point-source subtraction and host galaxy fitting for the 
red quasars. We did perform the subtraction for the most irregular merging 
systems in our sample (F2M0841$+$3604, F2M1113$+$1244, F2M1532$+$2415, 
F2M1656$+$3821); however, these sources have multiple components or are 
extremely irregular, so the host galaxy fitting will likely be wrong. For the 
fitting and subtracting we used our own IDL program {\it fithost}. It is 
based on the 2D fitting technique of \cite{mclure} and has been successfully 
applied to ground-based AO data \citep{lacy02}. The program simultaneously 
fits and subtracts both PSF and host galaxy; the host galaxy fitting is 
described in section \ref{host}.

\begin{figure*}
\setcounter{figure}{2}
\begin{center}
\vspace*{-0.5cm}
\includegraphics[width=12.3cm]{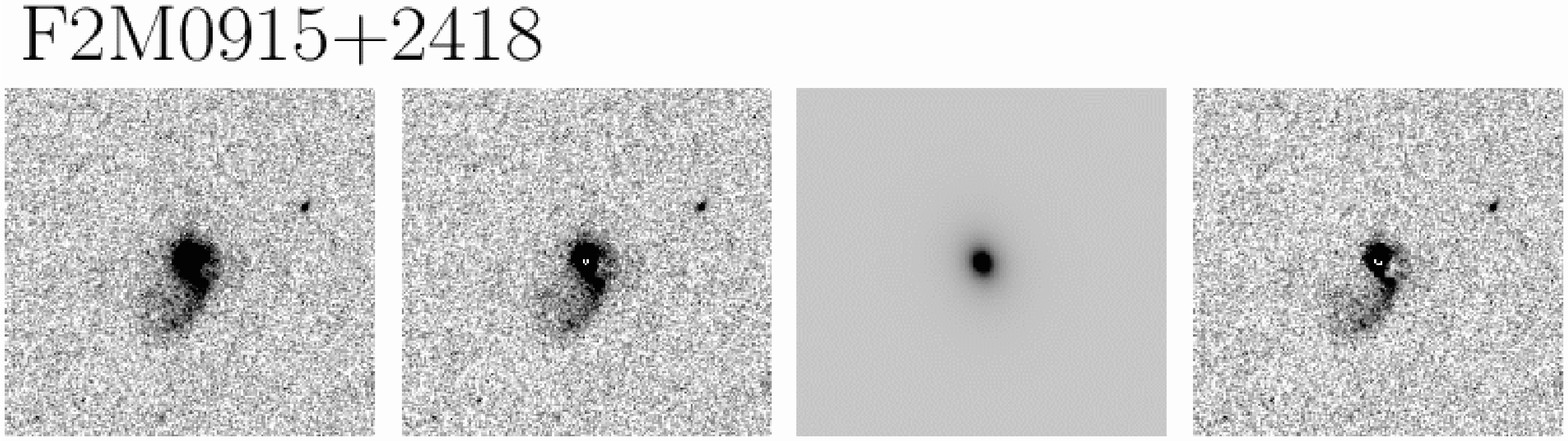}
\vspace*{0.5cm}
\includegraphics[width=4cm]{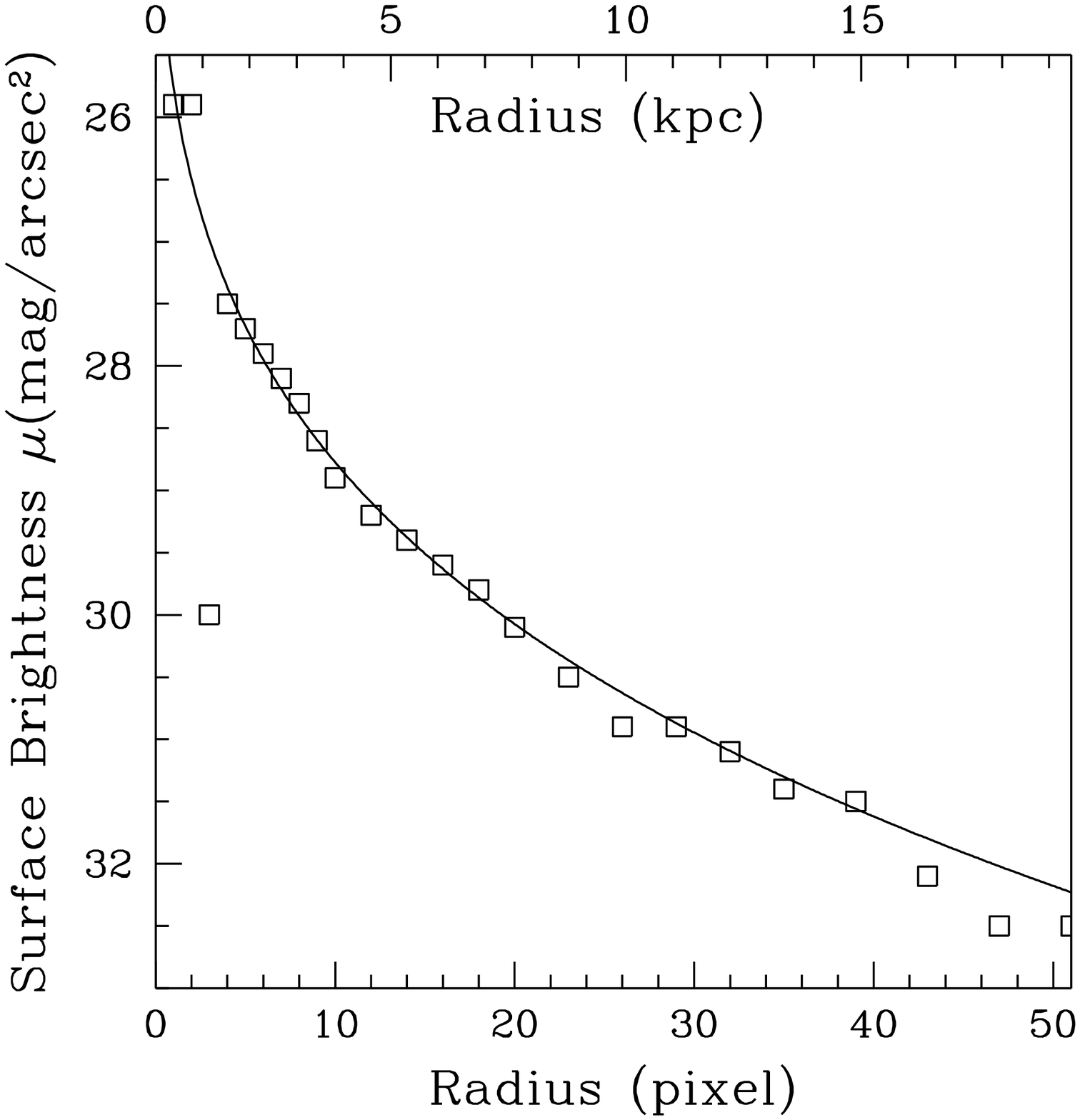}
\includegraphics[width=12.3cm]{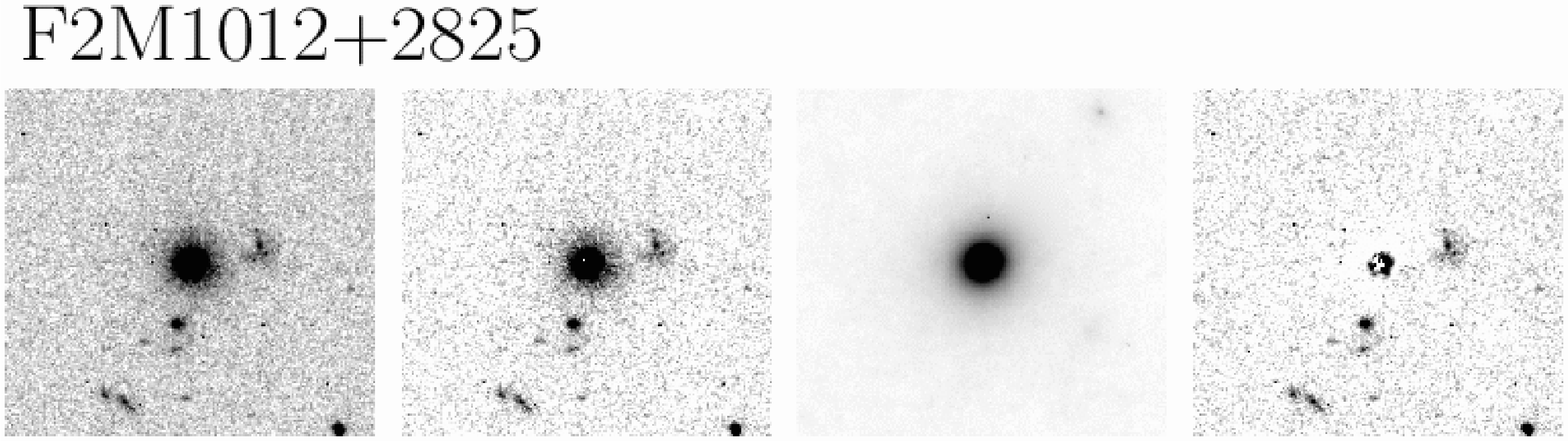}
\vspace*{0.5cm}
\includegraphics[width=4cm]{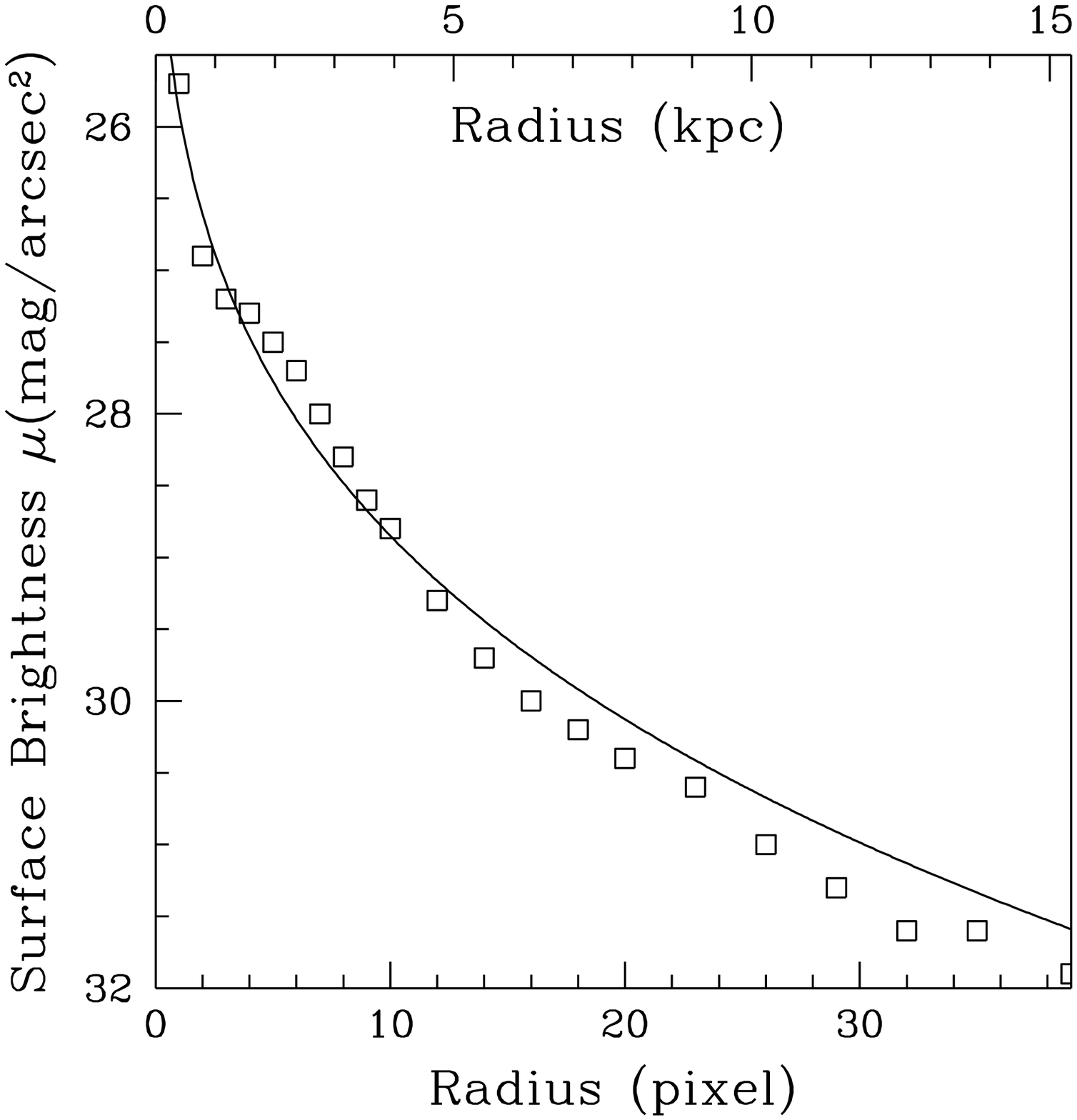}
\includegraphics[width=12.3cm]{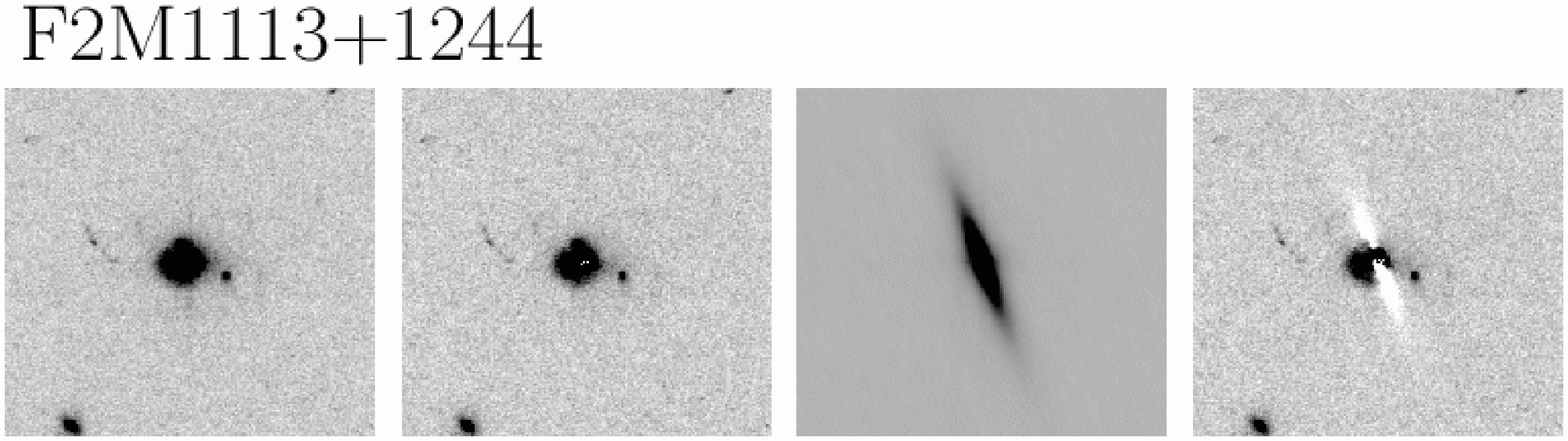}
\vspace*{0.5cm}
\includegraphics[width=4cm]{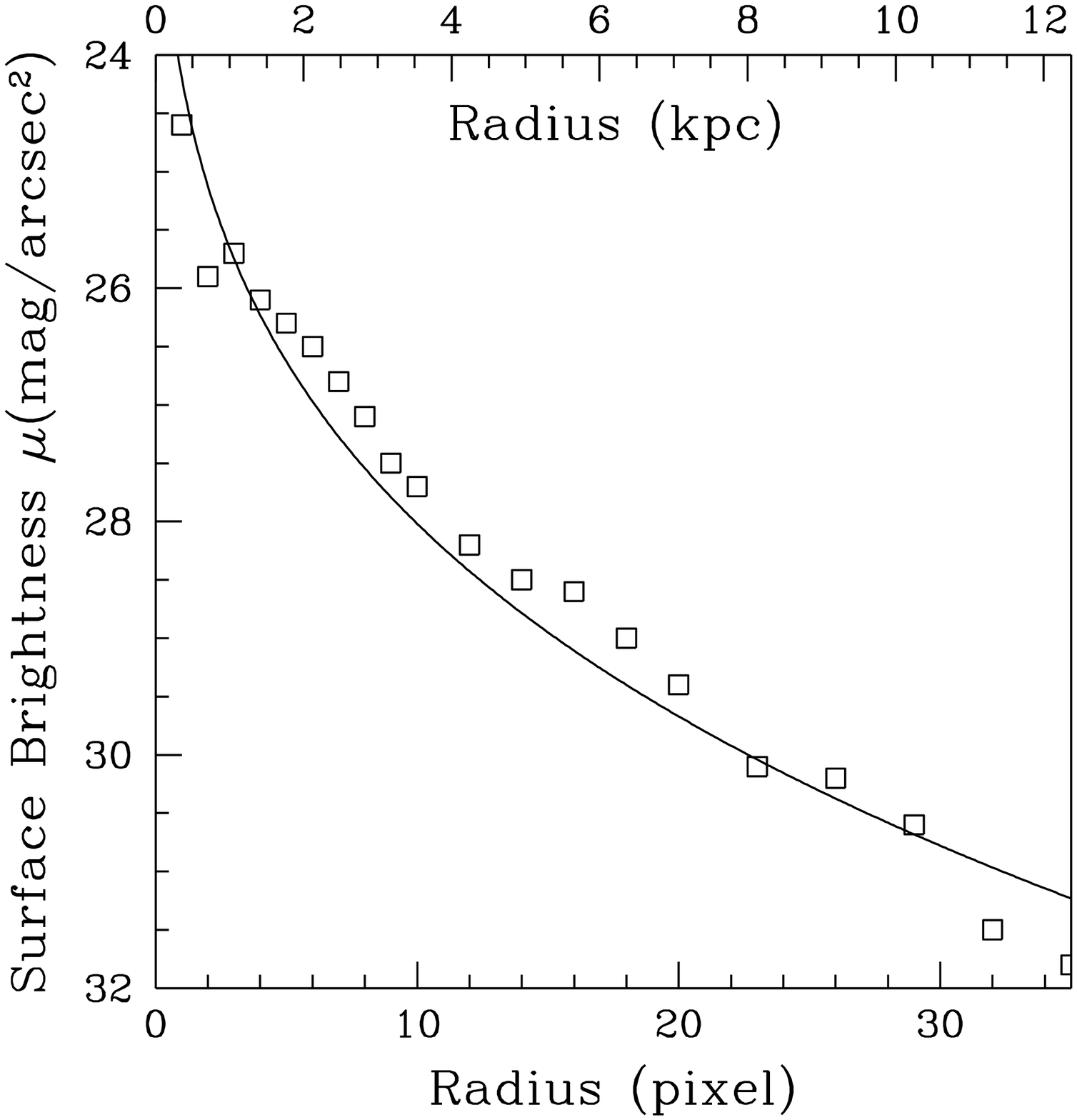}
\includegraphics[width=12.3cm]{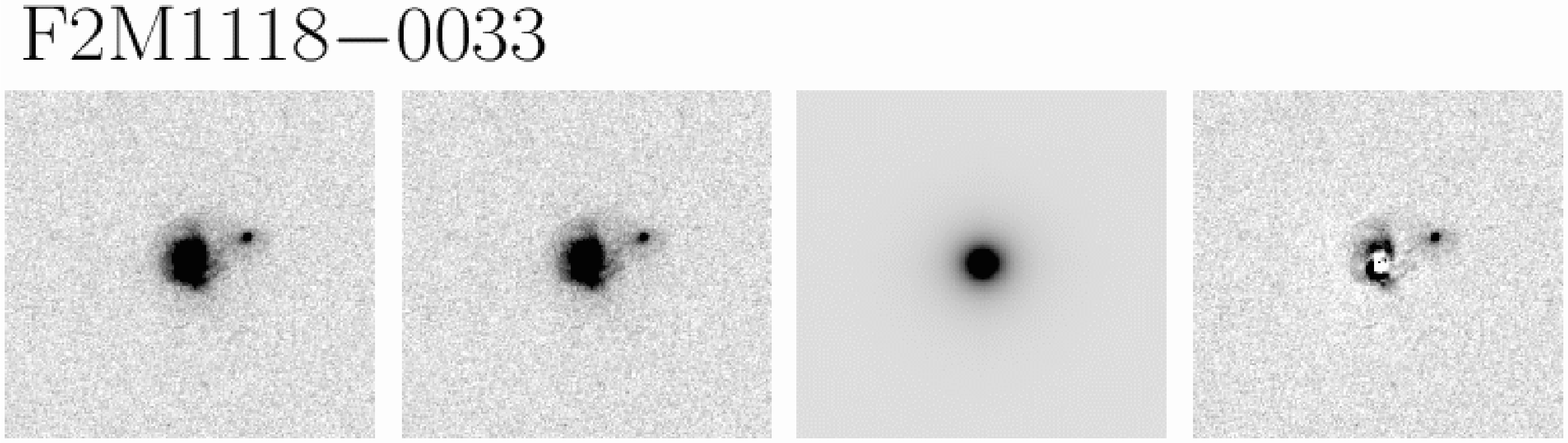}
\vspace*{0.5cm}
\includegraphics[width=4cm]{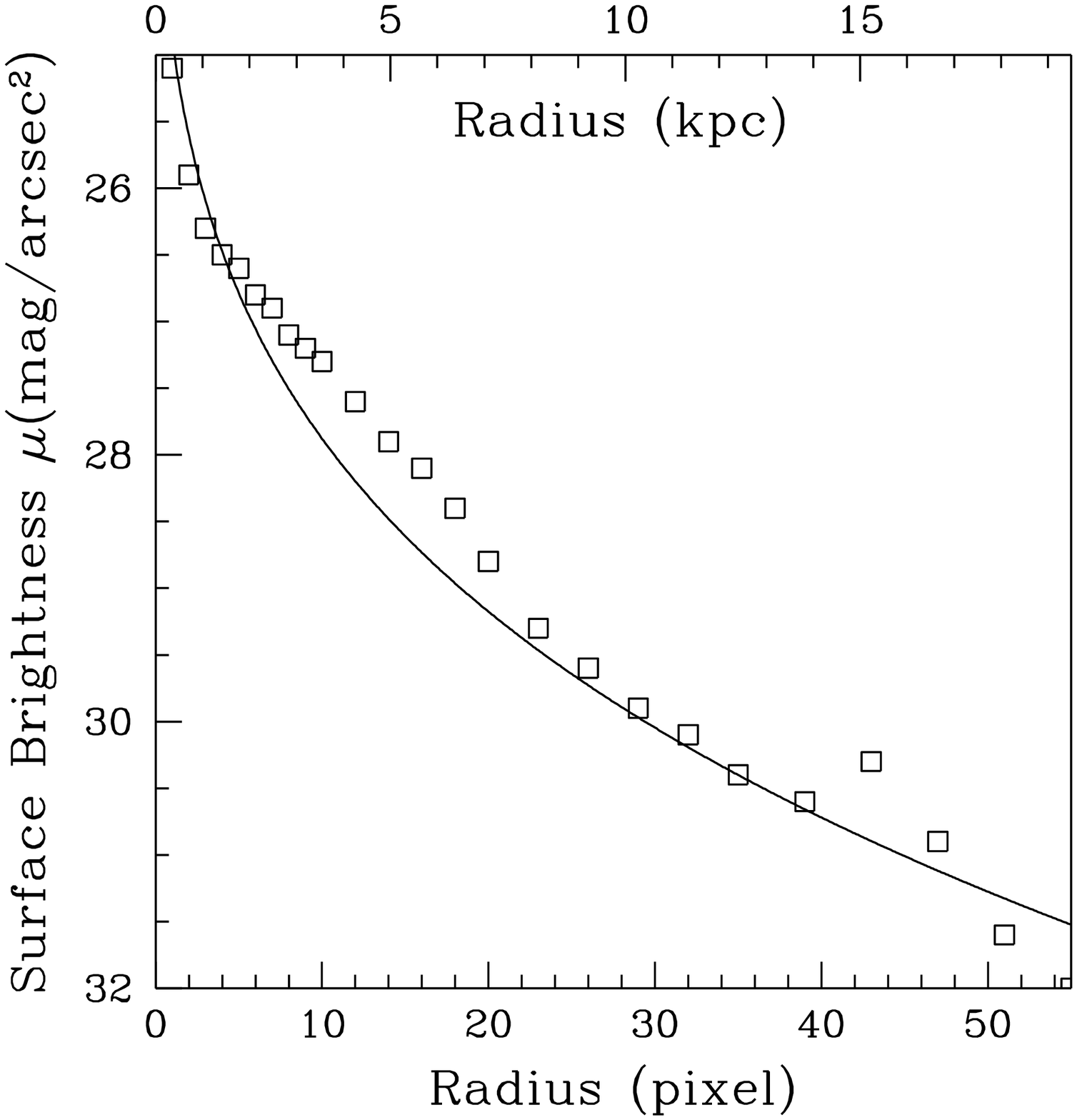}
\includegraphics[width=12.3cm]{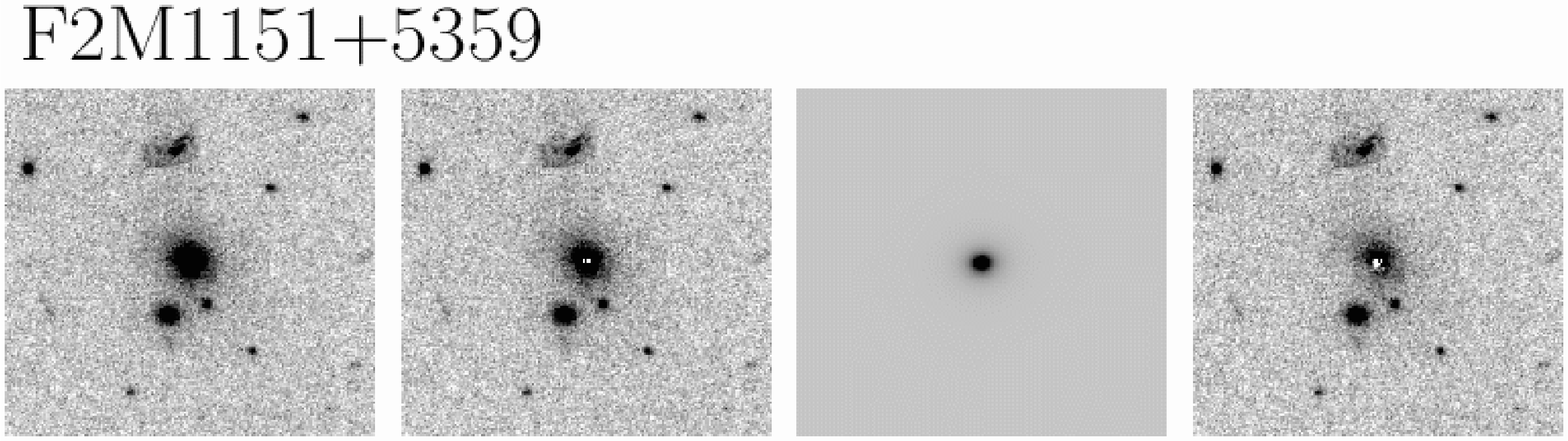}
\includegraphics[width=4cm]{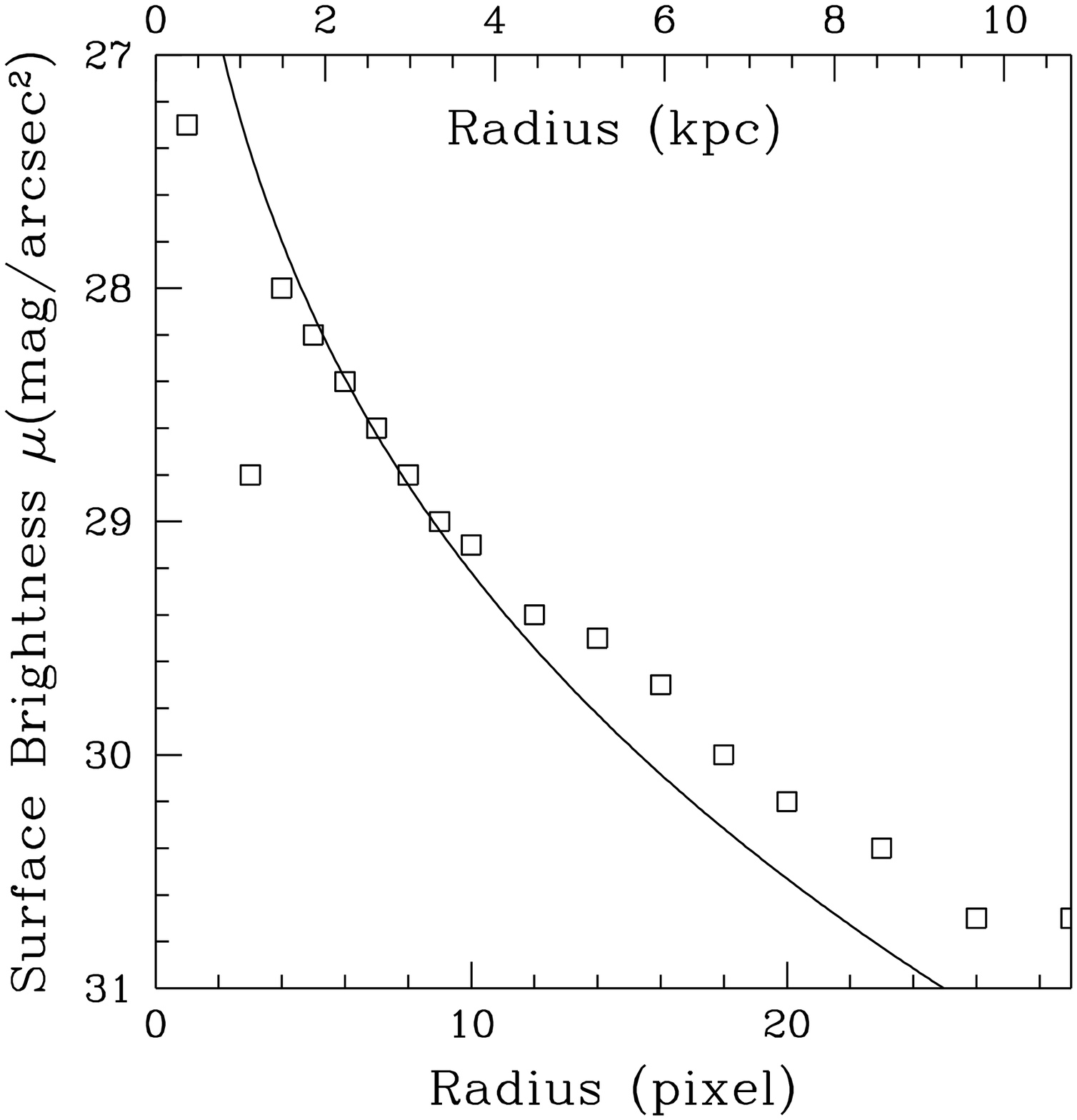}
\caption{cont.}
\end{center}
\end{figure*}

PSF stars were observed at the end of the HST orbits, immediately after the 
quasars. Stars were chosen to be close to the quasars, so as to avoid extra 
overhead due to guide star acquisition, and to be bright enough to have a 
similar fluence to the quasar in short exposures. Observations were made with 
the stars as close as possible to the same position as the quasars on the 
detector, and with the same dither offsets, The only drawback to this 
technique was that the ACS 1K subarray had to be used for both the quasars 
and the guide stars to avoid excessive readout overhead. Since we were mostly 
interested in the quasars and their close companions this was not a major 
problem.

\begin{figure*}
\setcounter{figure}{2}
\begin{center}
\vspace*{-0.5cm}
\includegraphics[width=12.3cm]{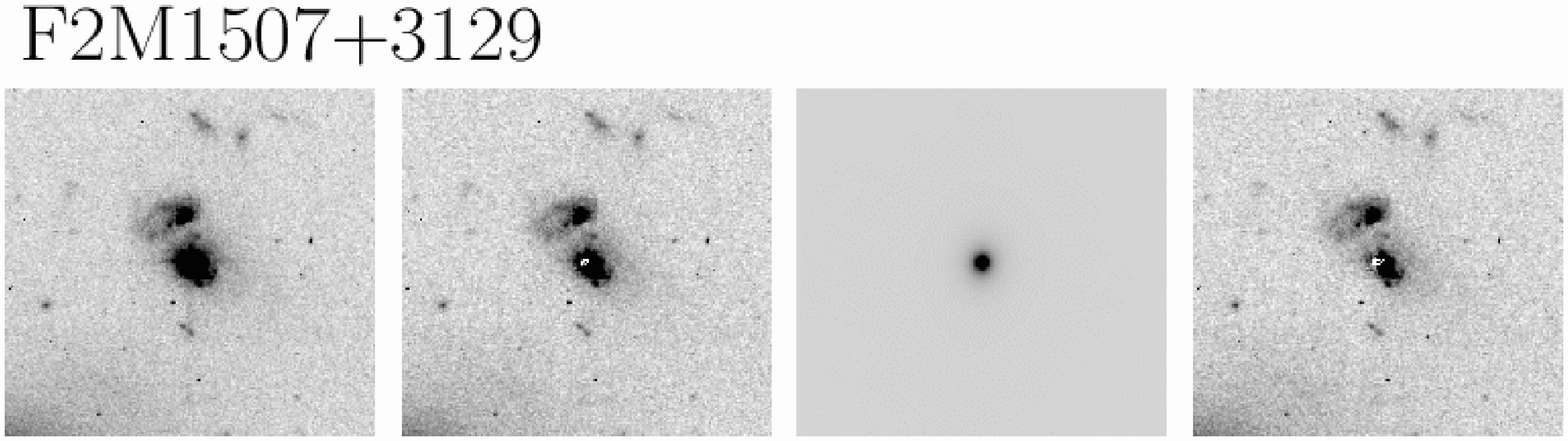}
\vspace*{0.5cm}
\includegraphics[width=4cm]{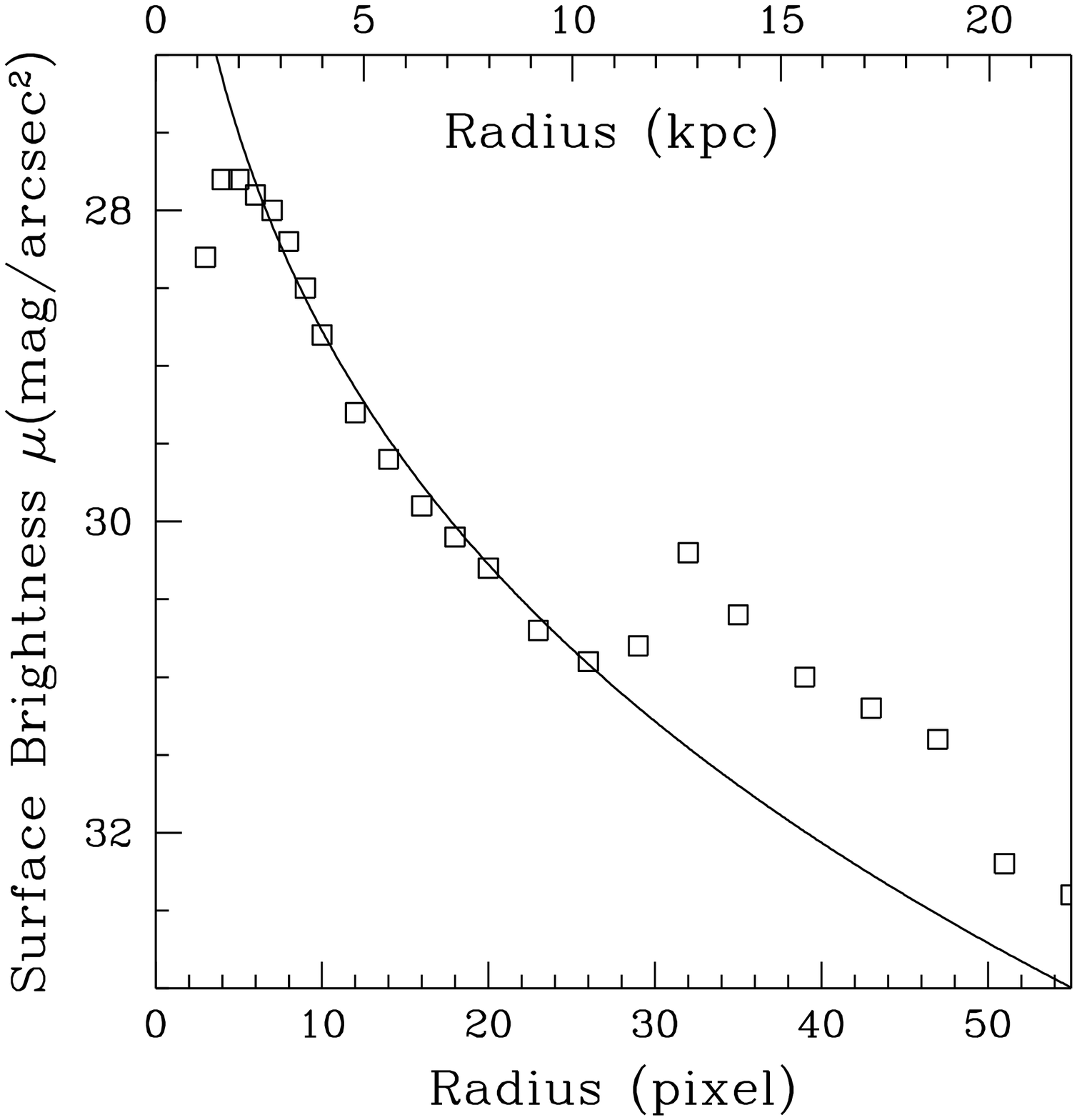}
\includegraphics[width=12.3cm]{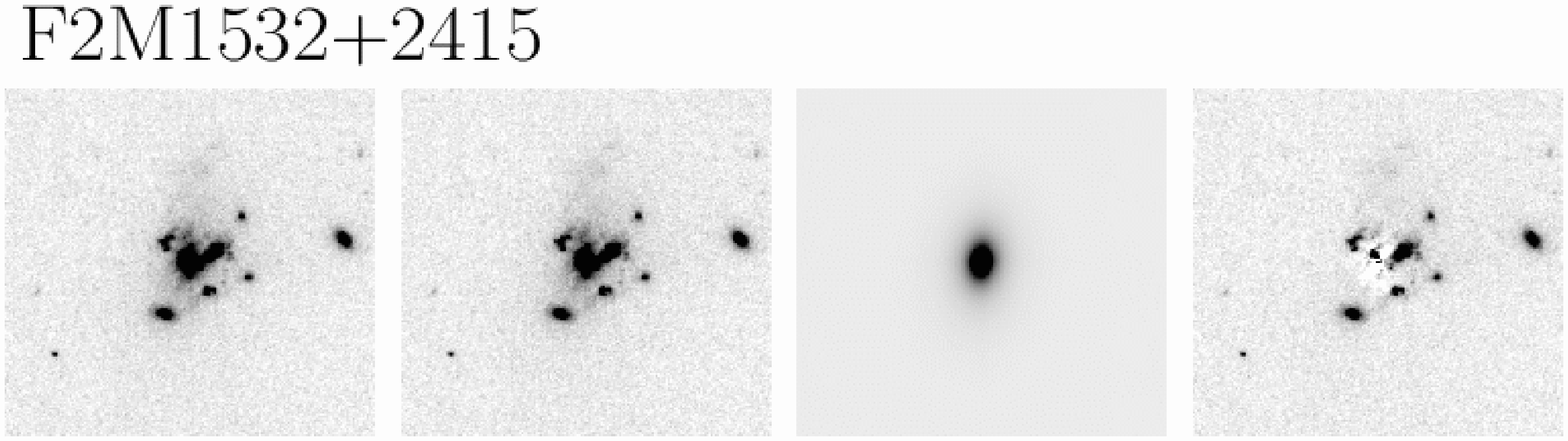}
\vspace*{0.5cm}
\includegraphics[width=4cm]{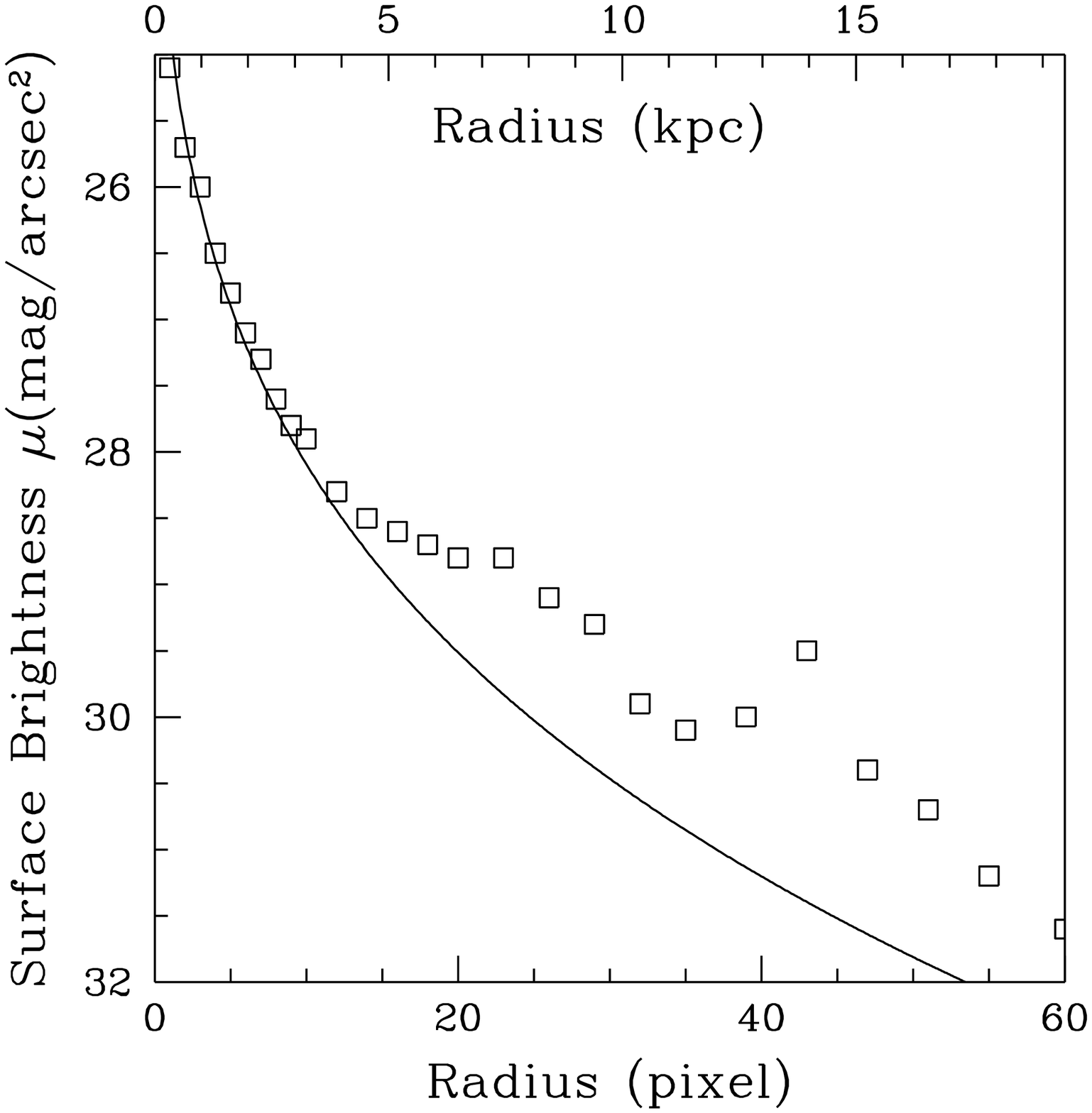}
\includegraphics[width=12.3cm]{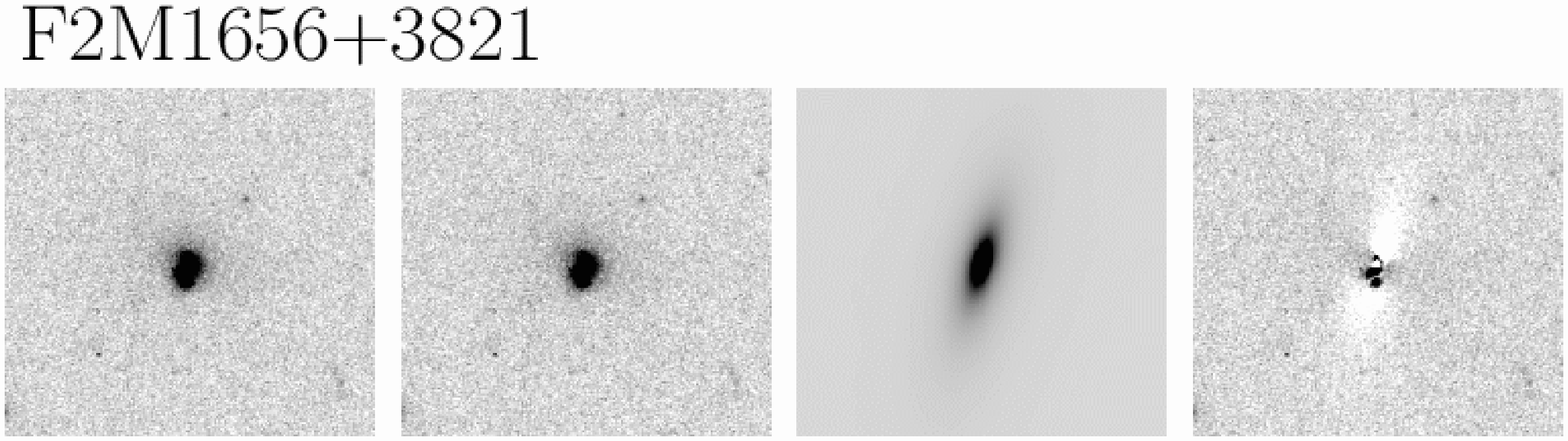}
\includegraphics[width=4cm]{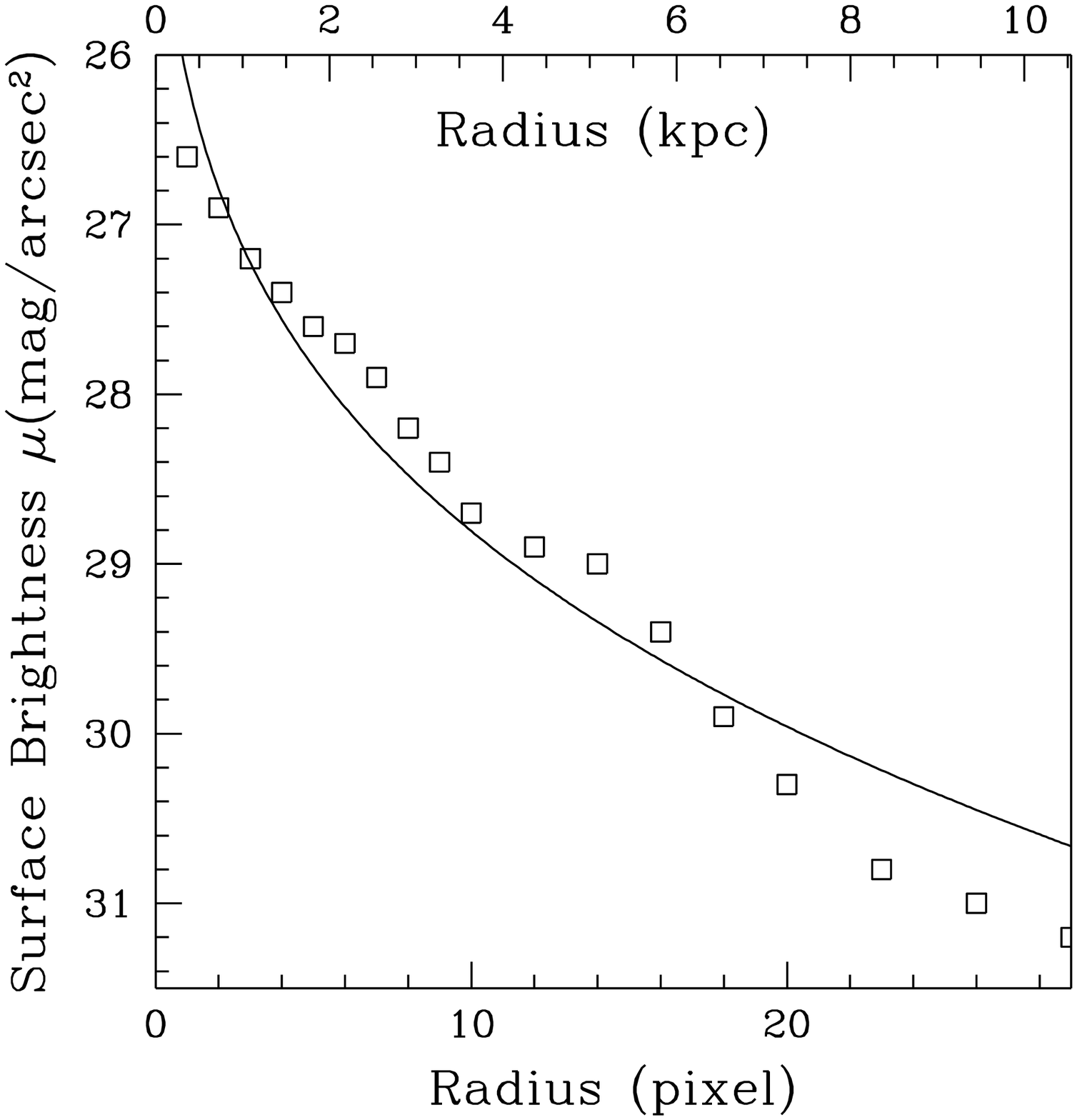}
\caption{cont.}
\end{center}
\end{figure*}

The {\it fithost} program was used in a mode where it automatically tries to 
fit the PSF-function to the brightest point in the image and then scale and 
remove it, so that the host galaxy can be subsequently fit. The center of the 
quasar and the host galaxy need not necessarily coincide. Starting parameters 
for the {\it fithost} program were derived using the IRAF program 
{\it imexam} and then refined with a second run. The first 2 postage stamp 
images in Figure \ref{subtraction} show the red quasars before and after 
PSF-subtraction. For the sake of brevity we chose to only show the $I_c$-band 
results, since they have most contribution from the quasar and most incident 
flux. In the $I_c$-band the flux from the PSF is typically similar to that of 
the host galaxies, but concentrated in the center of the image, which is why 
in Figure \ref{subtraction} we don't see glaring differences in the 
PSF-subtracted images from the original. In the $g^{'}$-band, the quasar is 
even fainter. The magnitude of the PSF/quasar after fitting is found in the 
first two columns of Table \ref{psf} 

The first result we can derive from the PSF fitting of the HST images is that 
the quasars are actually more obscured than deduced from the spectrum. In the 
HST $g^{'}$-band, which corresponds to rest-frame UV, we almost only see host 
galaxy contribution with the quasar being almost entirely extinguished. The 
$g^{'}$-$I_c$ colors the fitted quasar PSFs are also redder than for the total 
system. While the mean SDSS g'-i' color for the total system is 1.92, the 
average $g^{'}$-$I_c$ color for the quasar (just the PSF) is 0.61 magnitudes 
redder at 2.53.

We then calculated the reddening for only the quasar by reddening the slope 
of the FBQS composite (accounting for emission lines that might have fallen 
within the passbands) to fit the $g^{'}$-$I_c$ colors made by the PSF using 
the reddening curve described in section \ref{sample}. Most of the objects 
which already had large $E(B-V)$ reddenings around 1.0 increase their 
reddenings and colors, while the objects where the reddening of the total 
system were close to 0.5 on average don't increase or decrease their 
reddening. Interestingly, the objects for which the fit of the reddened 
quasar composite broke down, mostly decrease their reddenings, implying that 
the host galaxy itself has enough starlight that is responsible for some of 
the red colors of our objects. There could also be scattering of the nuclear 
light into the host galaxy itself, but we assume that the deduced PSF 
magnitude is the total quasar magnitude, since we can't assess the amount of 
scattering from the optical data alone. 

\begin{figure}
\begin{center}
\plotone{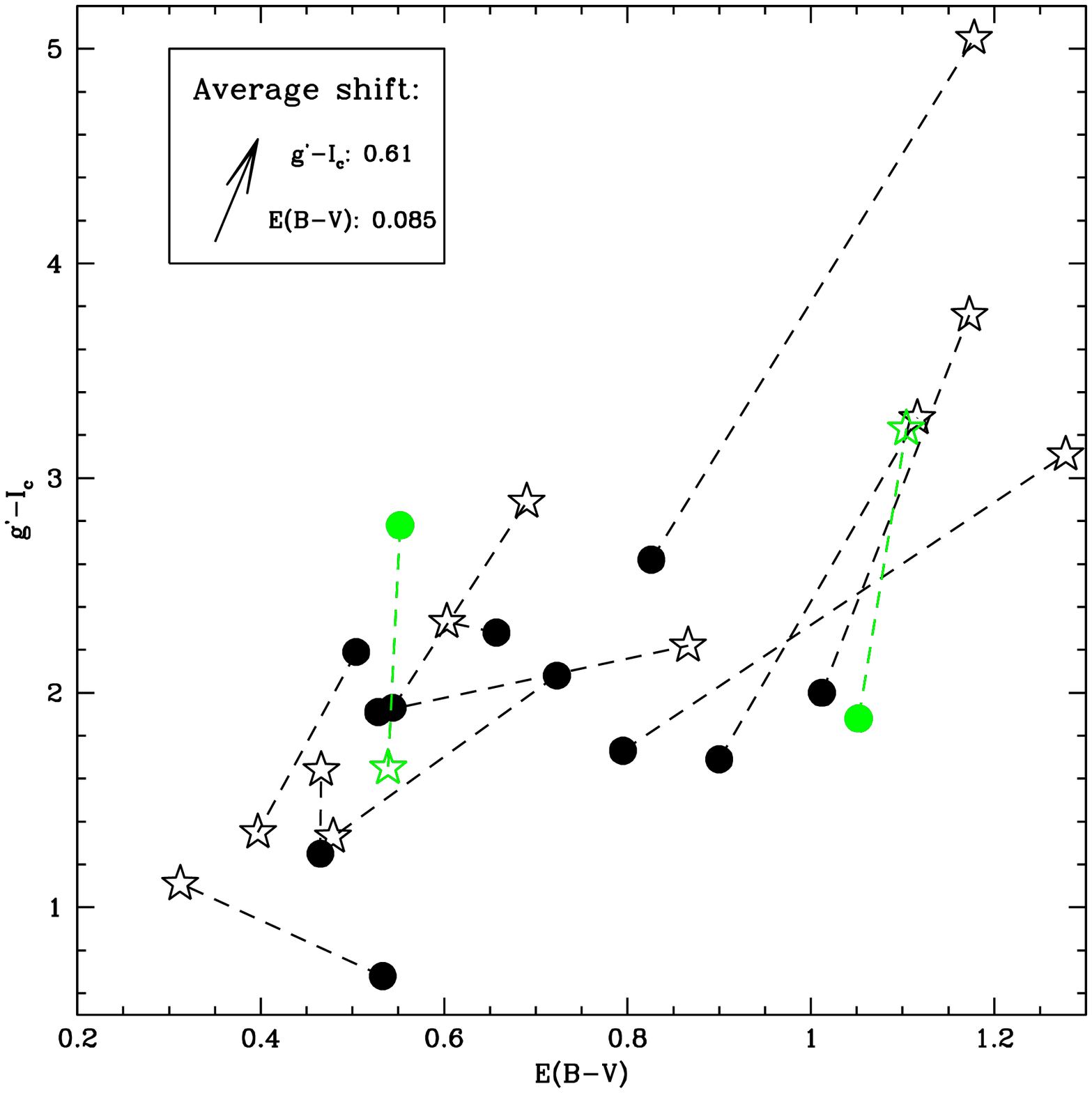}
\caption{$g^{'}$-$I_c$ color vs. $E(B-V)$ reddening. The filled dots 
represent the magnitudes and reddenings for the total system. The magnitudes 
are from SDSS data and don't differ much from HST photometry and the $E(B-V)$ 
reddening is from spectral fitting. The open stars represent quasar (PSF) 
color and reddenings derived from HST PSF photometry. We included lines to 
show how the values shifted. On average the quasar has a 0.61 mag redder 
color and higher $E(B-V)$ of 0.085 mag.}
\label{giebv}
\end{center}
\end{figure}

Figure \ref{giebv} shows the shifts of the objects colors and reddenings. The 
total magnitudes and reddenings are represented by the filled circles and the 
stars represent the quasar/PSF. Two of the quasars (F2M0841$+$3604 and 
F2M1656$+$3821) were not resolved in the SDSS, so the PSF magnitudes are much 
fainter than implied by SDSS. We mark this objects in Figure \ref{giebv} in 
green color, their shifts are likely to be wrong, but we include them in the 
Figure for completeness.

From the $g^{'}$-$I_c$ colors, magnitudes and reddenings, we can then 
calculate the true luminosity for the quasar, since the absolute magnitudes 
calculated in section \ref{sample} from the spectrum had contamination from 
the host galaxy, especially in the bluer passband. Table \ref{psf} gives the 
new luminosities (corrected for reddening) determined from the quasar colors 
and magnitudes. The quasars are a bit more luminous on average, due to the 
fact that the quasar is more obscured than had been deduced from the 
spectrum. The quasars on average are still well above the quasar/Seyfert 
divide and much more luminous than the red quasar study conducted by 
\cite{marble}, so our claim that we are missing a large population of 
obscured AGN is still valid.

\begin{center}
\begin{deluxetable*}{ccccccccc}
\tabletypesize\scriptsize
\tablecaption{Quasar (PSF) properties}
\tablewidth{0pt}
\tablehead{ \multicolumn{1}{c|}{Source} & 
\multicolumn{2}{|c|}{PSF magnitudes} & \multicolumn{2}{|c|}{g-I colors} & 
\multicolumn{2}{|c|}{$E(B-V)$} & \multicolumn{2}{|c|}{Luminosities ($M_B$)} \\
\multicolumn{1}{c|}{} & \multicolumn{1}{|c}{Mag$_{I_c}$} & 
\multicolumn{1}{c|}{Mag$_{g^{'}}$} & \multicolumn{1}{|c}{Total} & 
\multicolumn{1}{c|}{PSF} & \multicolumn{1}{|c}{Total} & 
\multicolumn{1}{c|}{PSF} & \multicolumn{1}{|c}{Total} & 
\multicolumn{1}{c|}{PSF} \\
\multicolumn{1}{c|}{} & \multicolumn{2}{|c|}{} &  
\multicolumn{1}{|c}{(SDSS)} & \multicolumn{1}{c|}{(HST)} & 
\multicolumn{1}{|c}{(spectrum)} & \multicolumn{1}{c|}{(HST)} &
\multicolumn{1}{|c}{spectrum} & \multicolumn{1}{c|}{HST}}
\startdata
F2M0729$+$3336 & 20.15 $\pm$ 0.07 & 25.20 $\pm$ 0.73 & 2.62$^a$ & 5.05 & 
0.83 $\pm$ 0.22 & 1.18 $\pm$ 0.16 & $-$24.96 & $-$26.83 \\
F2M0825$+$4716 & 22.33 $\pm$ 0.21 & 23.65 $\pm$ 0.36 & 2.19 & 1.32 & 
0.50 $\pm$ 0.35 & 0.40 $\pm$ 0.14 & $-$21.94 & $-$21.92 \\
F2M0830$+$3759 & 19.51 $\pm$ 0.06 & 22.62 $\pm$ 0.22 & 1.73 & 3.11 & 
0.80 $\pm$ 0.15 & 1.29 $\pm$ 0.12 & $-$22.73 & $-$23.07 \\
F2M0834$+$3506 & 18.95 $\pm$ 0.04 & 21.17 $\pm$ 0.11 & 1.91 & 2.22 & 
0.53 $\pm$ 0.10 & 0.87 $\pm$ 0.06 & $-$22.90 & $-$23.80 \\
F2M0841$+$3604 & 21.88 $\pm$ 0.17 & 25.11 $\pm$ 0.70 & 1.88 & 3.23 & 
1.05 $\pm$ 0.64 & 1.10 $\pm$ 0.27 & $-$21.03 & $-$21.61$^b$ \\
F2M0915$+$2418 & 19.88 $\pm$ 0.07 & 20.99 $\pm$ 0.10 & 0.68 & 1.11 & 
0.53 $\pm$ 0.36 & 0.31 $\pm$ 0.04 & $-$23.69 & $-$24.43 \\
F2M1012$+$2825 & 21.17 $\pm$ 0.12 & 23.50 $\pm$ 0.33 & 2.28 & 2.33 & 
0.66 $\pm$ 0.12 & 0.60 $\pm$ 0.12 & $-$24.37 & $-$24.26$^b$ \\
F2M1113$+$1244 & 19.11 $\pm$ 0.05 & 22.87 $\pm$ 0.25 & 2.00 & 3.76 & 
1.01 $\pm$ 0.24 & 1.17 $\pm$ 0.09 & $-$25.08 & $-$25.45 \\
F2M1118$-$0033 & 22.00 $\pm$ 0.18 & 23.33 $\pm$ 0.32 & 2.08 & 1.33 & 
0.72 $\pm$ 0.26 & 0.48 $\pm$ 0.15 & $-$22.97 & $-$21.76 \\
F2M1151$+$5359 & 20.48 $\pm$ 0.09 & 22.12 $\pm$ 0.18 & 1.25 & 1.59 & 
0.47 $\pm$ 0.13 & 0.47 $\pm$ 0.07 & $-$23.31 & $-$23.78 \\
F2M1507$+$3129 & 19.78 $\pm$ 0.07 & 22.67 $\pm$ 0.23 & 1.93 & 2.89 & 
0.54 $\pm$ 0.13 & 0.69 $\pm$ 0.08 & $-$24.27 & $-$26.24 \\
F2M1532$+$2415 & 21.14 $\pm$ 0.12 & 24.42 $\pm$ 0.51 & 1.69 & 3.28 & 
0.90 $\pm$ 0.53 & 1.12 $\pm$ 0.21 & $-$20.46 & $-$22.45$^b$ \\
F2M1656$+$3821 & 23.46 $\pm$ 0.35 & 25.11 $\pm$ 0.70 & 2.78 & 1.65 & 
0.55 $\pm$ 0.57 & 0.54 $\pm$ 0.28 & $-$20.86 & $-$20.66$^b$ \\
\enddata
\tablecomments{$^a$No SDSS photometry, we use HST values. 
$^b$ Luminosity is only for one nucleus. \label{psf}}
\end{deluxetable*}
\end{center}

\subsection{Properties of the Host Galaxies}\label{host}

Before modelling the full host/quasar system, we obtained a constraint on the 
host galaxy magnitude. We did this by constraining the scaled PSF subtraction 
so that the residual was approximately flat in the center of the quasar host 
within a 3 pixel radius and declined monotonically outside that radius. We 
also obtained the lower limit of the host galaxy magnitude by subtracting the 
PSF until the central pixel was zero. The ``mono'' and ``zero'' magnitude are 
found in Table \ref{galaxy}. 

The host galaxies were modeled by fitting PSF plus galaxy model profiles 
(convolved by the PSF) by minimizing $\chi^2$. Close to the center of the 
source, systematic errors from the PSF subtraction dominate, so to prevent 
those errors to dominate the fit, the inner 3 pixel radius was downweighted 
by a factor of 0.5 to ensure that the $\chi^2$ surface was fairly uniform 
across the fitting aperture. The position of the PSF and the galaxy nucleus 
was not held fixed; so in total we fit the flux, angle, axial ratio and 
position of the host galaxy to the image. We tried to fit both elliptical and 
exponential profiles to the images, however in all, but one case 
(F2M1118$-$0033), the elliptical profile was best, so all our quoted values 
are for an elliptical profile.

Figure \ref{subtraction} shows the fitted elliptical galaxy and the residuals 
after PSF and model-galaxy subtraction in the third and fourth panels 
respectively. Note that none of the galaxies are a perfect fit to an 
elliptical, however, there were some which came quite close, like 
F2M0915$+$2418. Also, two of the host galaxies, while they did not fit the 
elliptical profile well, did not show signs for interaction (F2M0834$+$3506 
and F2M1151$+$5359). The fraction of 11 out of 13 systems showing interaction 
is much higher than the 2MASS red AGN fraction of one third \citep{marble} 
and other unobscured quasar host galaxy studies. The fifth panel in Figure 
\ref{subtraction} shows the radial surface brightness profiles for the PSF 
subtracted images and the solid line is the model applied with which they 
were fitted.

In the following paragraphs, when we discuss the host galaxies we mean the 
PSF-subtracted images and not the modeled host galaxies, to assess their 
magnitudes, morphologies and comparison the the quasar.

\begin{figure*}
\begin{center}
\includegraphics[width=4.8cm]{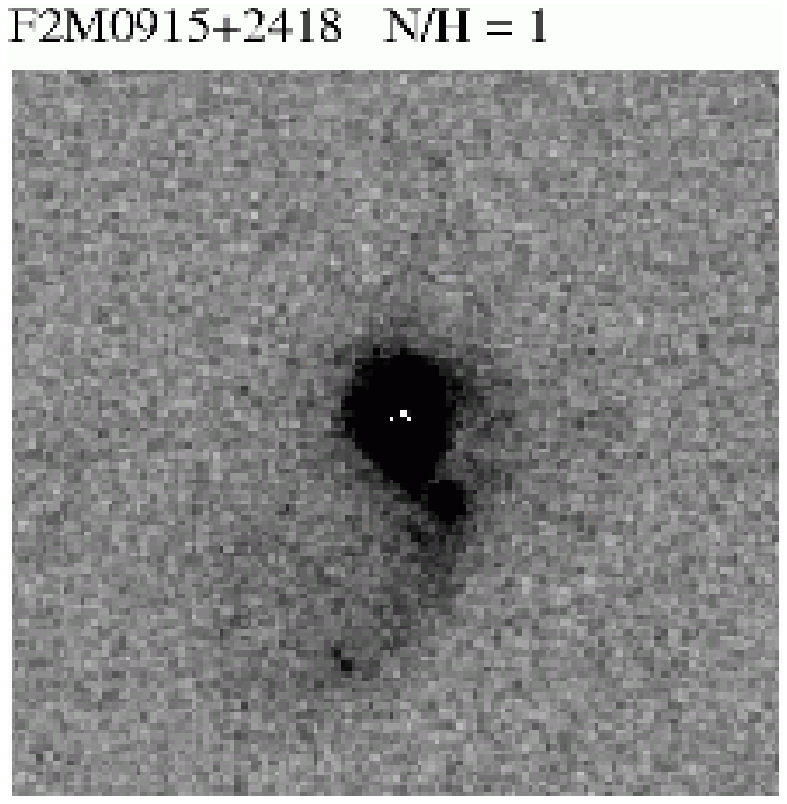}
\hspace*{0.4cm}
\vspace*{0.4cm}
\includegraphics[width=4.8cm]{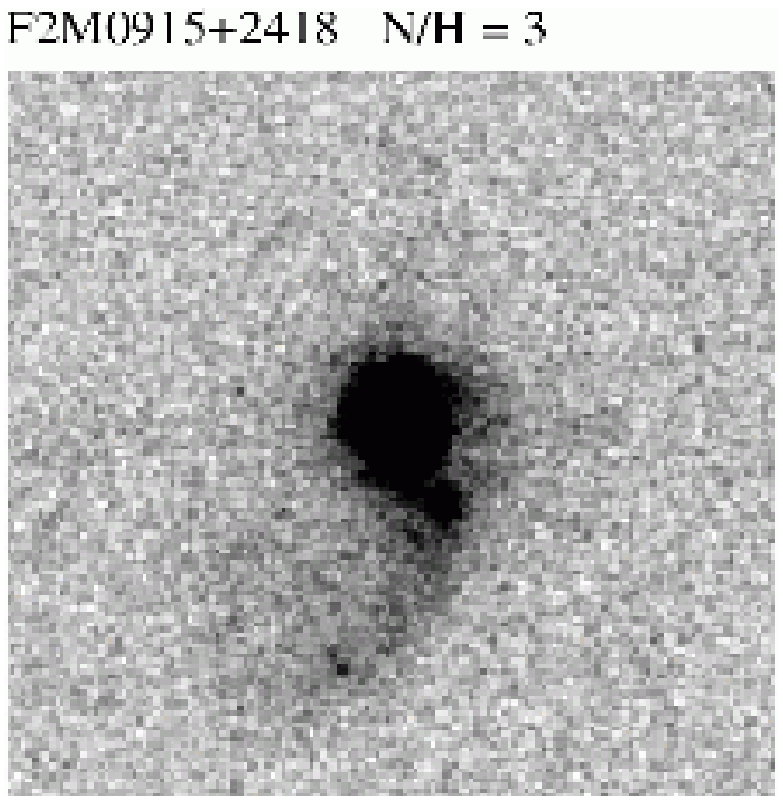}
\hspace*{0.4cm}
\includegraphics[width=4.8cm]{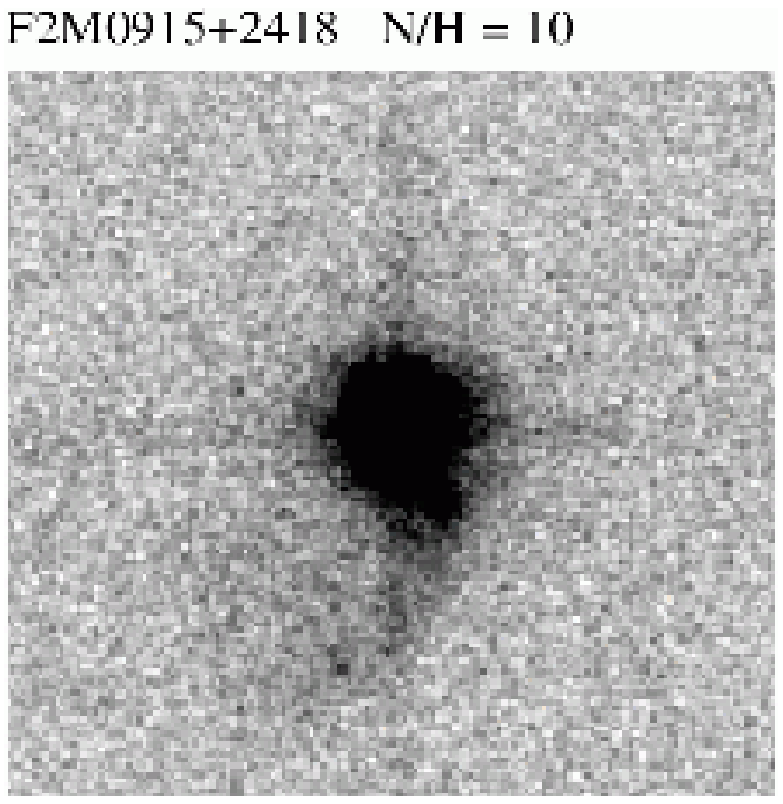}
\includegraphics[width=4.8cm]{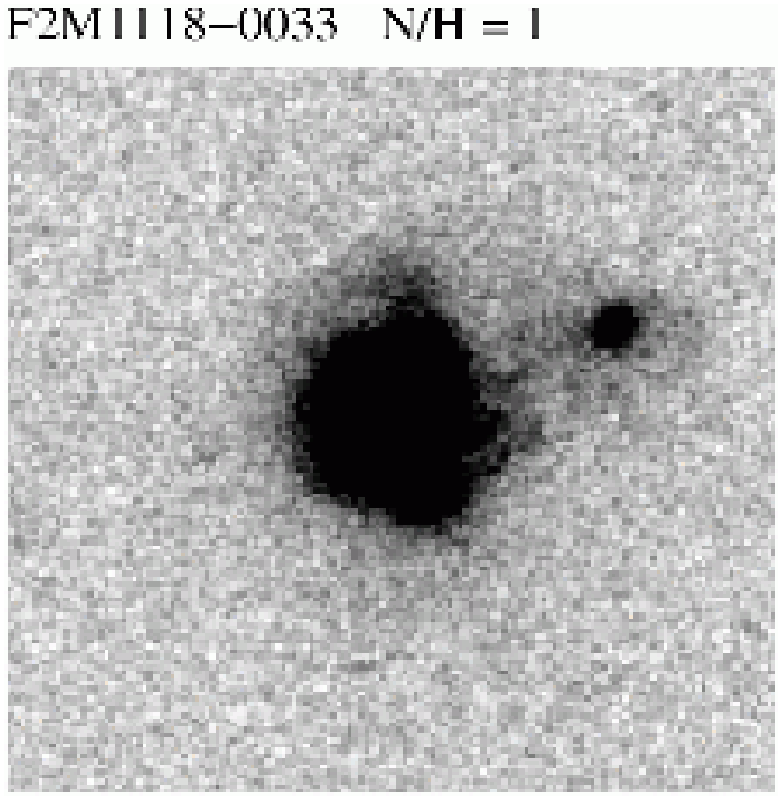}
\hspace*{0.4cm}
\vspace*{0.4cm}
\includegraphics[width=4.8cm]{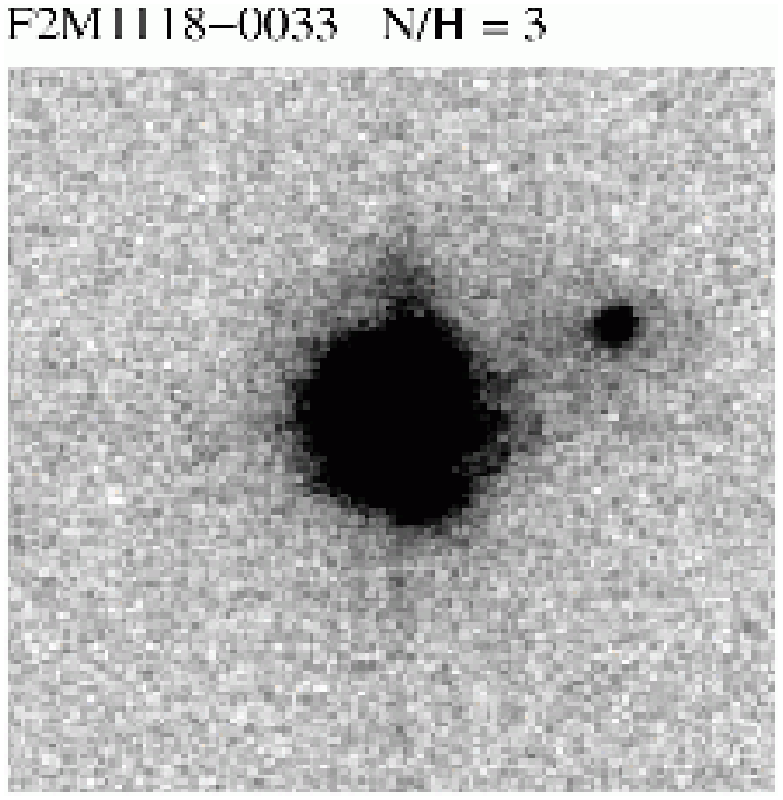}
\hspace*{0.4cm}
\includegraphics[width=4.8cm]{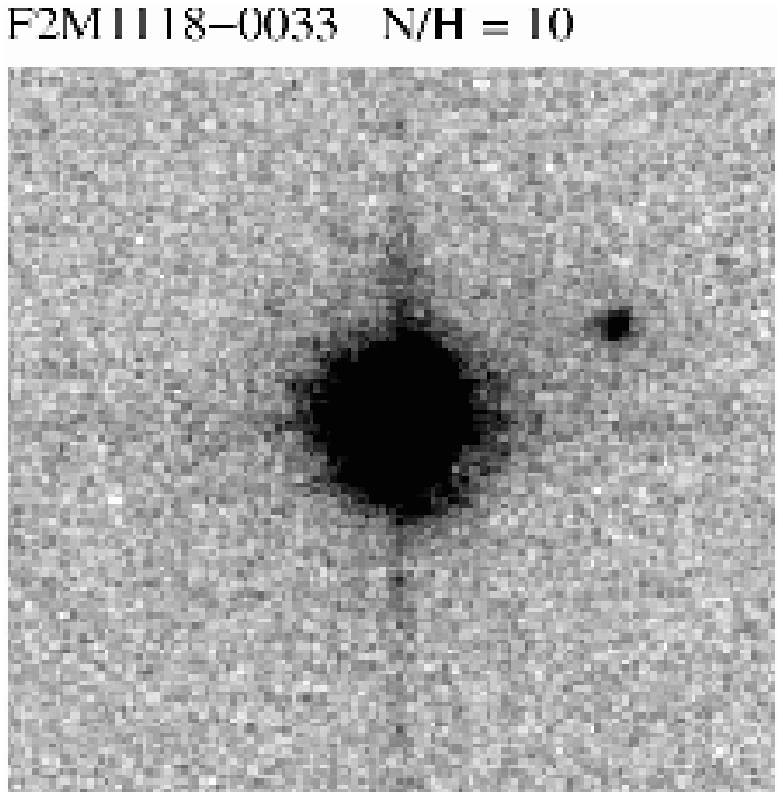}
\includegraphics[width=4.8cm]{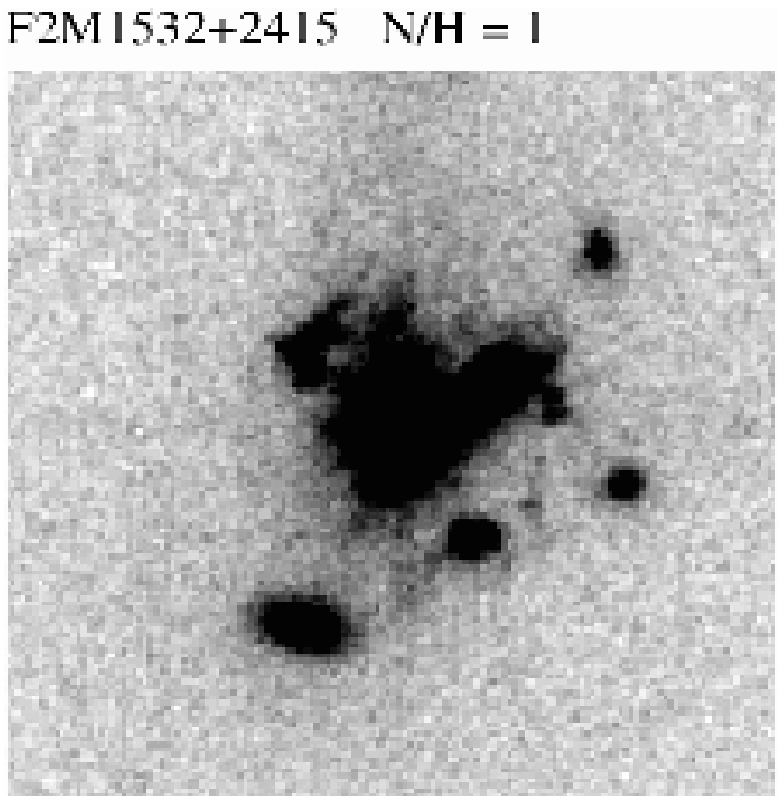}
\hspace*{0.4cm}
\includegraphics[width=4.8cm]{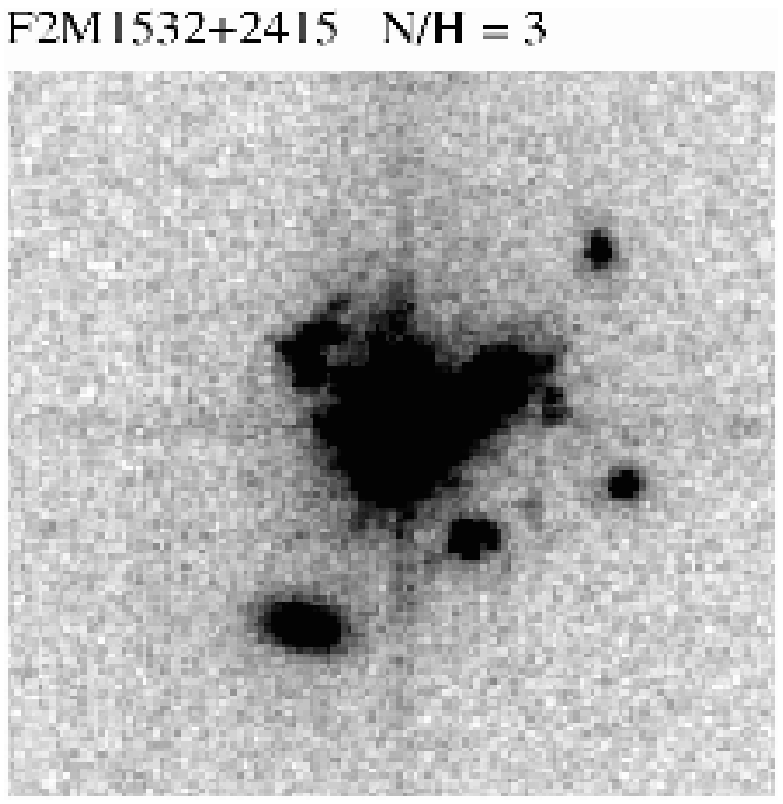}
\hspace*{0.4cm}
\includegraphics[width=4.8cm]{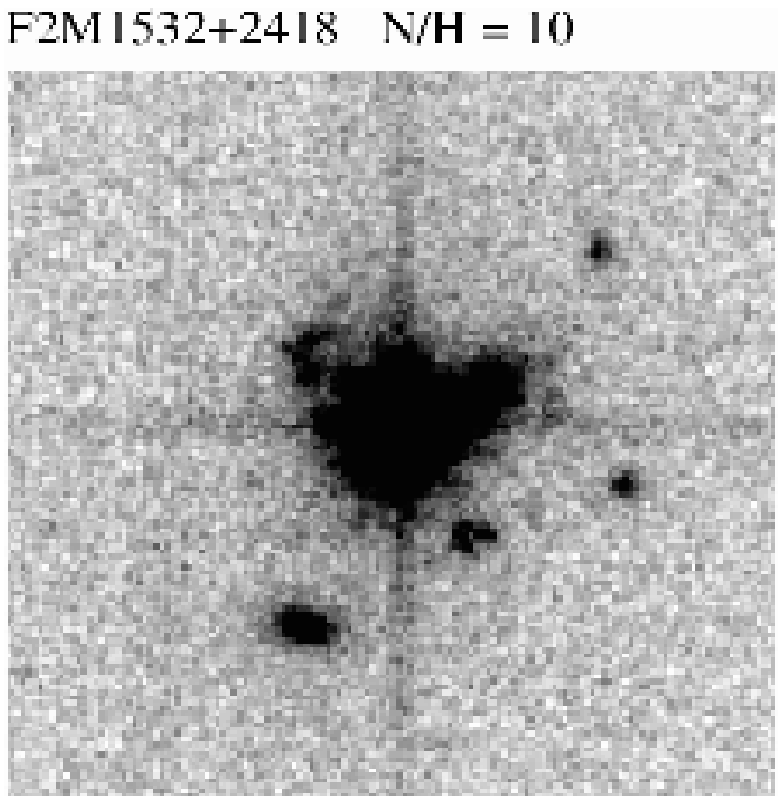}
\caption{Host galaxies with artificial PSF added. The images show different 
nucleus to host ratios in increasing order. When the nucleus has the same 
magnitude as the host galaxy, which is the case for the red quasars in this 
sample, the interactions are clearly visible, such as the tidal connection in 
F2M1118$-$0033 or the strong halo in F2M1532$+$2415. With a nucleus to host 
ratio of 10, most of the features have disappeared and the PSF clearly 
dominates, as is typical in unobscured quasar sampled. \label{nhfig}}
\end{center}
\end{figure*}

We measured the magnitudes of the host galaxies with Sextractor following the 
conventions above. The isophotal-magnitudes are quoted in Table \ref{galaxy}. 
They were calculated in the same way the magnitudes were calculated in 
Section \ref{acs}.  Note that in the cases where the total magnitude of the 
host galaxy is fainter than the limit denoted by the ``zero'' magnitude, the 
total host magnitude only refers to the central component of the galaxy, as 
the host galaxies often showed so large interactions that they are broken up 
and have various components. 

The colors of the host galaxies are not particularly blue, so the host 
galaxies do not show large amounts of unobscured star formation. However, 
some of the extended components are blue, indicating the presence of either 
merger-induced star formation, or light associated with the quasar. This 
quasar light could be either scattered continuum, or extended line emission. 
The lack of a clear ionization cone morphology (with the possible exception of
F2M0830$+$3759), however, would suggest that most of this blue light is from 
young stars.

We also measured the nucleus to host (N/H) ratios from these magnitudes 
(Table \ref{galaxy} for the $I_c$-band N/H). As discussed in section 
\ref{propqso}, the N/H for the $g^{'}$-band will be lower, because the quasar 
is more extinguished and most of the light in that band will come from the 
host. Overall, the luminosity of the hosts are on average, a little larger 
than those of the nuclei, such as the sample of IR-excess quasar sample of 
\cite{surace}. The relatively low N/H are in sharp contrast to the high N/H 
ratios of the luminous quasar sample of \cite{mclure99}, which is tied to the 
\cite{dunlop} sample. However the N/H ratios of our quasars are generally 
higher than the red quasar sample of \cite{marble}. 

We then calculated the absolute magnitudes ($M_B$) of the hosts adopting a 1 
Gyr post starburst model from \cite{bc} for the K-correction. At the typical 
redshifts of our sample (z $\sim$ 0.7) $M^*_B = -20.30$ \citep{glf}, so the 
host galaxies are around or a bit above $L^*$ luminosity.

One measure of the extent of the host galaxy is the Petrosian radius $r_p$ 
\citep{petrosian}. Very irregular or interacting systems tend to have high 
Petrosian radii, while compact ellipticals have a low $r_p$. The Petrosian 
radius is defined when the ratio of he surface brightness at the Petrosian 
radius and the average surface brightness at radii below reach a certain 
value $\eta$:

\begin{equation}
\eta = \frac{\mu(r_p)}{\bar{\mu}(r < r_p)}	
\end{equation}

Following SDSS conventions, we set $\eta = 0.2$. Table \ref{galaxy} quotes 
the Petrosian radii of our systems. With one exception (the very compact 
F2M0729$+$3336), most have Petrosian radii around 1''-2.5'', which are normal 
for galaxies at those redshifts. When comparing those numbers to the surface 
density plots in Figure \ref{subtraction}, one can notice that the nucleus 
($r < r_p$) of the host galaxy often has an elliptical profile; only the most 
irregular systems such as F2M0841$+$3604 are irregular at low radii. Mostly, 
the interaction features only appear well beyond the Petrosian radius and 
usually have low surface brightness. 

This could be an indication why many authors conclude that the host galaxies 
of luminous quasars are fit best mostly by elliptical profiles (e.g. 
\citep{floyd}), but find so few merger remnants. Perhaps with the 
overpowering AGN (high N/H ratios), the low surface brightness features or 
interaction morphology signatures are lost. Recently, very deep imaging of 
some of the quasars studied by \cite{dunlop}, show either low surface 
brightness tidal tails or other merger remnants in the form of shells 
\citep{nicola} further supporting this point. The red quasars in this study 
are quite extinguished, so the host galaxy features are easier to discern. We 
tested how the tidal features can disappear in the presence of a bright 
quasar by adding point sources of different brightness (1,3,10 N/H ratios) to 
the host galaxy image. When a PSF with a N/H ratio of 10 is added, the quasar 
is so bright that in all cases with exception of F2M0841$+$3604 the red 
quasars lose most of their interaction features and the quasar dominates the 
image (Figure \ref{nhfig}). Even so, after performing PSF subtraction on a 
N/H=10 quasar+host system, the interactions should reappear. Also the red 
quasar host galaxies are in an earlier stage of the merger as we will see next.

As we already commented in section \ref{acs} the fraction of our red quasars 
showing interaction in their host galaxies is very high. Furthermore, all of 
the {\it dust} reddened quasars, that is the quasars where the dust reddening 
template fit best, show interaction. E(B-V) seems to correlate weakly with 
the amount of interaction; the more obscured the quasar is, the more 
disturbed the morphology of the host galaxy is. We can arrive at these 
conclusions more or less ``by eye'', but we support our claims by 
parameterizing the host galaxy morphology with their Gini coefficients and 
their Concentration indices.

\begin{center}
\begin{deluxetable*}{ccccccccccccc}
\tabletypesize\scriptsize
\tablecaption{Host galaxy properties}
\tablewidth{0pt}
\tablehead{ \multicolumn{1}{c|}{Source} & 
\multicolumn{3}{|c|}{Host mag$_{I_c}$} & 
\multicolumn{3}{|c|}{Host mag$_{g^{'}}$} & 
\multicolumn{3}{|c|}{Luminosities} & 
\multicolumn{3}{|c|}{Morphologies} \\
\multicolumn{1}{c|}{} & \multicolumn{1}{|c}{zero} & \colhead{mono} & 
\multicolumn{1}{c|}{total} & \multicolumn{1}{|c}{zero} & \colhead{mono} & 
\multicolumn{1}{c}{total} & \multicolumn{1}{|c}{N/H$^a$} & \colhead{$M_B$} &
\multicolumn{1}{c|}{L$^*$} & \multicolumn{1}{|c}{$r_p$ (pix)} & 
\colhead{G} & \multicolumn{1}{c|}{C}}
\startdata
F2M0729$+$3336 & 20.71 & 19.97 & 19.78 $\pm$ 0.06 & 21.80 & 21.80 & 21.78 
$\pm$ 0.15 & 0.711 & $-$22.32 & 6.4 & 03.8 & 0.69 & 0.36 \\
F2M0825$+$4716 & 20.07 & 19.99 & 21.48 $\pm$ 0.14 & 21.71 & 21.58 & 22.95 
$\pm$ 0.26 & 0.470 & $-$20.02 & 0.8 &29.3 & 0.54 & 0.42 \\
F2M0830$+$3759 & 18.96 & 18.75 & 18.86 $\pm$ 0.04 & 20.53 & 20.22 & 20.31 
$\pm$ 0.08 & 0.550 & $-$21.02 & 1.9 &25.9 & 0.57 & 0.41 \\
F2M0834$+$3506 & 19.96 & 19.87 & 19.77 $\pm$ 0.06 & 22.66 & 22.56 & 21.13 
$\pm$ 0.11 & 2.089 & $-$20.42 & 1.1 &37.5 & 0.52 & 0.55 \\
F2M0841$+$3604 & 19.96 & 19.90 & 20.40 $\pm$ 0.09 & 21.57 & 21.54 & 23.04 
$\pm$ 0.27 & 0.256 & $-$20.22 & 0.9 &53.7 & 0.52 & 0.30 \\
F2M0915$+$2418 & 20.72 & 20.64 & 20.82 $\pm$ 0.10 & 25.58 & 23.83 & 22.91 
$\pm$ 0.25 & 2.355 & $-$20.90 & 1.7 &24.0 & 0.49 & 0.38 \\
F2M1012$+$2825 & 20.56 & 20.44 & 20.87 $\pm$ 0.11 & 22.12 & 21.99 & 23.49 
$\pm$ 0.33 & 0.759 & $-$21.15 & 2.2 &19.8 & 0.59 & 0.52 \\
F2M1113$+$1244 & 20.06 & 19.76 & 19.75 $\pm$ 0.06 & 21.12 & 21.87 & 21.31 
$\pm$ 0.12 & 1.803 & $-$21.28 & 2.5 &18.6 & 0.68 & 0.55 \\
F2M1118$-$0033 & 19.34 & 19.24 & 19.51 $\pm$ 0.06 & 20.95 & 20.91 & 22.70 
$\pm$ 0.22 & 0.101 & $-$21.53 & 3.1 &22.6 & 0.59 & 0.42 \\
F2M1151$+$5359 & 20.03 & 19.99 & 20.99 $\pm$ 0.11 & 22.70 & 22.46 & 22.43 
$\pm$ 0.20 & 1.528 & $-$20.41 & 1.1 &49.6 & 0.39 & 0.41 \\
F2M1507$+$3129 & 20.10 & 20.07 & 20.96 $\pm$ 0.11 & 22.24 & 22.09 & 22.45 
$\pm$ 0.21 & 2.858 & $-$21.18 & 2.2 &48.5 & 0.57 & 0.49 \\
F2M1532$+$2415 & 19.15 & 19.05 & 19.39 $\pm$ 0.05 & 20.77 & 20.69 & 22.20 
$\pm$ 0.18 & 0.194 & $-$21.27 & 2.4 &33.4 & 0.60 & 0.40 \\
F2M1656$+$3821 & 20.58 & 20.45 & 20.77 $\pm$ 0.10 & 22.22 & 22.14 & 24.96 
$\pm$ 0.62 & 0.039 & $-$20.47 & 1.1 &20.3 & 0.59 & 0.44 \\
\enddata
\tablecomments{$^a$Nucleus to host ratios from $I_c$-Band magnitudes.\\
Zero and monoton are model magnitudes, therefore they don't 
have errors.\label{galaxy}}
\end{deluxetable*}
\end{center}

The Gini coefficient is a non-parametric approach to classifying a galaxy, 
and can therefore be easily applied to irregular galaxies or highly merging 
systems. It is a measure of the cumulative distribution of a galaxy's pixel 
values and is a good alternative approach to quantifying the amount of 
interactions in galaxies. The Gini coefficient is correlated with the 
concentration index of galaxies for spiral and elliptical galaxies 
\citep{abraham03} and anti-correlated with the Concentration Index for ULIRGs 
and other highly irregular systems \citep{lotz}. A high Gini coefficient 
therefore indicates either a high level of interactions or a highly 
concentrated elliptical galaxy, while a low Gini coefficient indicates either 
spiral galaxies or low surface brightness galaxies.

We calculated the Gini coefficient of the host galaxies of the red quasars by 
following the conventions in \cite{abraham03}. Based on the formalism by 
\cite{glasser}, if we sort the pixels $X_i$'s flux into increasing order, the 
Gini coefficient can be calculated by:

\begin{equation}
G = \frac{1}{\bar{X} n (n-1)} \sum_i^n (2i - n - 1) X_i 
\end{equation}

However, we took notice of the warning by \cite{lotz} and tried to create 
segmentation maps at the $\mu(r_p)$, the flux threshold above which pixels 
are assigned to the galaxy. For that we cut out galaxy postage stamps cutouts 
by eye and included only pixels within that cutout that had a surface 
brightness higher than $\mu(r_p)$. The Gini coefficient was then computed of 
the distribution of {\it absolute} flux values of the pixels which corrects 
for the areas in the center which were PSF-oversubtracted. The Gini 
coefficients G can be found in Table \ref{galaxy}.

The Concentration Index measures the ratio of the flux withing an inner 
radius to that within and outer circular or elliptical aperture. The main 
difference in the definitions by different authors is the choice of the radii 
or semimajor axes of the two apertures. \cite{conselice} adopts a ratio 
between the radii containing certain percentages of the total light. 
\cite{abraham94} uses a flux ratio between two normalized radii $\alpha$ and 
1, where E(1) is the area encompassing 2$\sigma$ flux and $\alpha$ is 
typically set to 0.3. 

\begin{equation}\label{conindex}
C = \frac{\sum \sum_{i,j \in E(\alpha)} I_{ij}}
{\sum \sum_{i,j \in E(1)} I_{ij}}
\end{equation}

Since \cite{abraham03} has already linked the Gini coefficient to the 
Concentration index via a unity slope, we decided to use the Concentration 
index defined in equation \ref{conindex} instead of the \cite{conselice} 
definition. We chose the total flux as the flux within 1.5$r_p$ and 
E($\alpha$) as the flux within 0.45$r_p$. From that ratio we are able to 
obtain C, which is quoted in Table \ref{galaxy}.

\begin{figure*}
\begin{center}
\includegraphics[width=7cm]{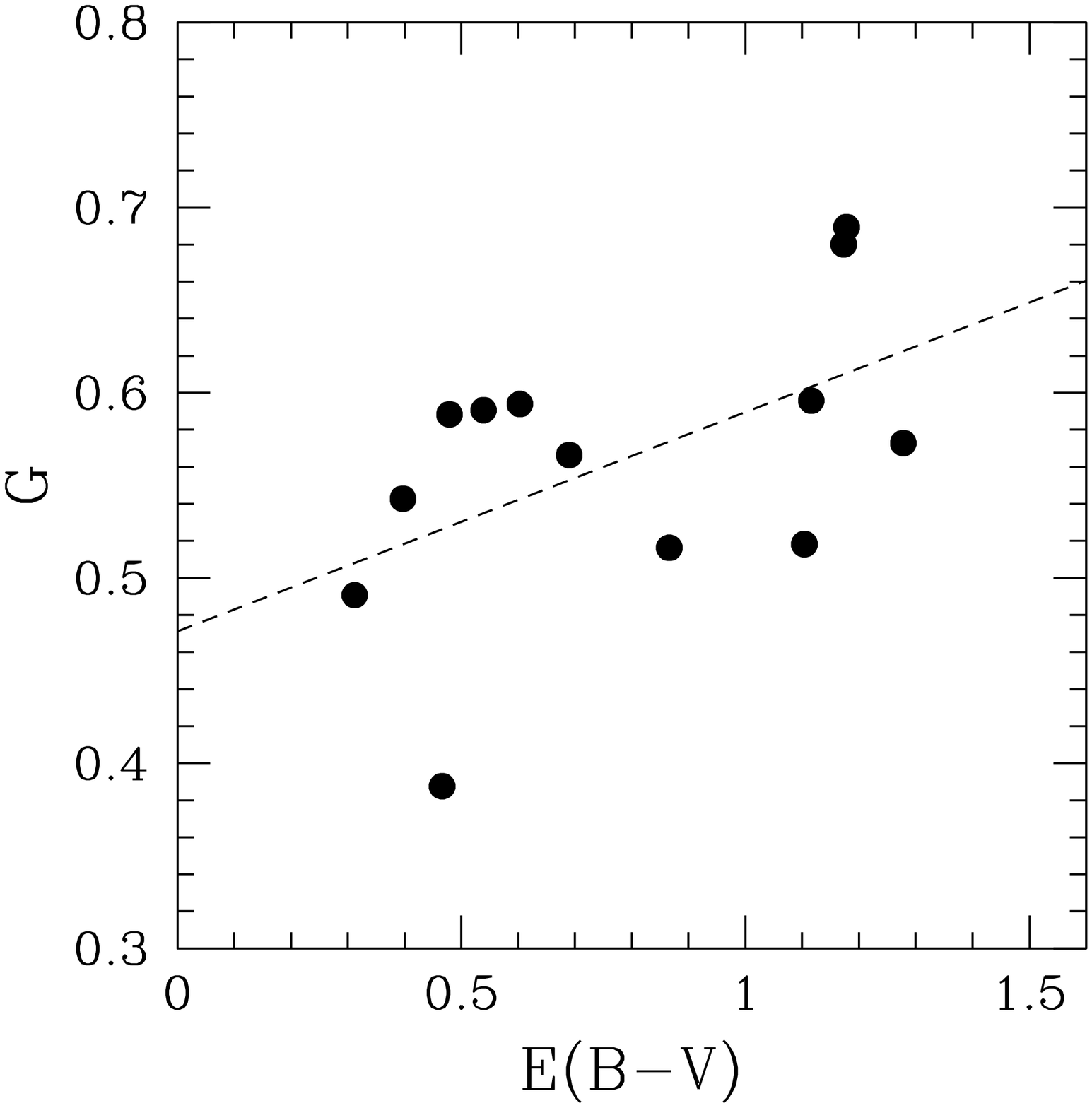}
\hspace*{0.2cm}
\includegraphics[width=7cm]{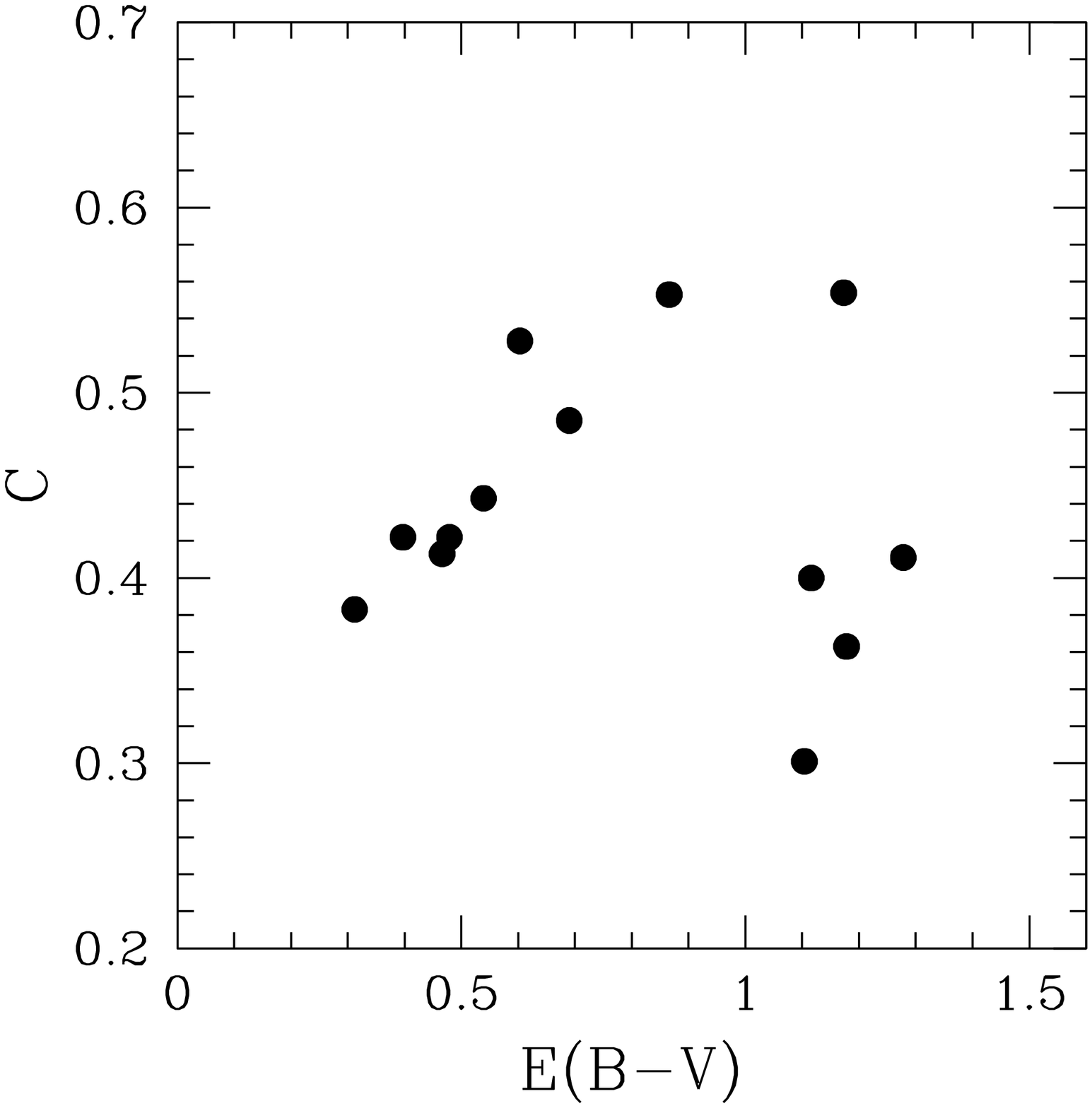}
\includegraphics[width=7cm]{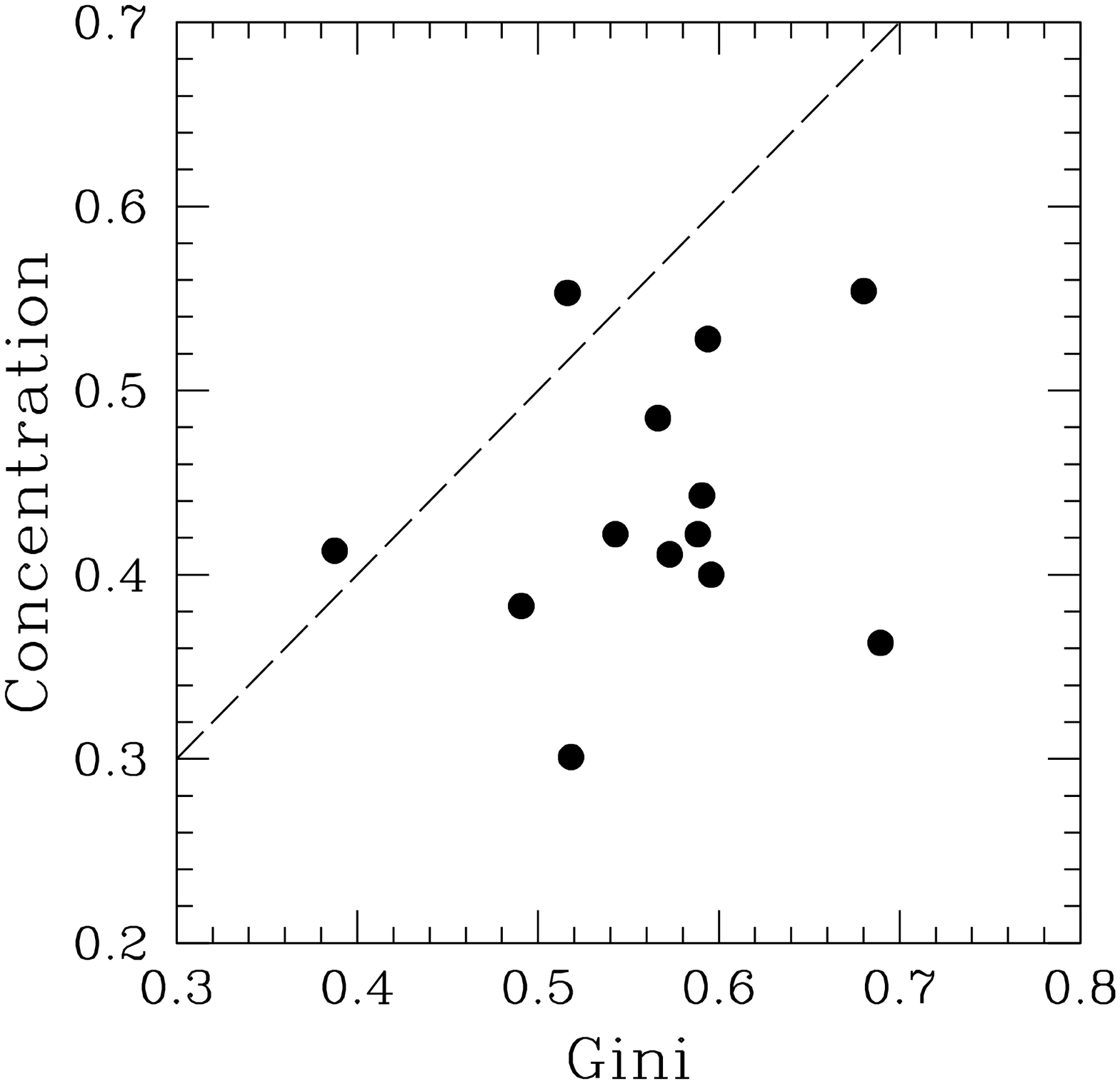}
\caption{Upper left graphic: Gini coefficient vs. reddening. There is a weak 
correlation between the Gini coefficient and the reddening (dashed line). 
High Gini coefficient indicate either highly irregular systems, such as 
ULIRGs or very compact ellipticals. Upper right graphic: Concentration index 
vs. reddening. For single systems the Gini coefficient and the Concentration 
index are correlated. The fact that for high reddening we don't necessarily 
have high Concentration indices means that the systems are highly disturbed, 
usually with more than one nucleus. Lower graphic: Concentration index vs. 
Gini coefficient. No correlation is found in our sample in contrast to other 
studies of single galaxies, indicating that a many of the red quasar hosts 
are in a merger status. The dotted line is the unity correlation found by 
\cite{abraham03}. All but two host galaxies (the two undisturbed ones) of the 
F2M red quasars have higher Gini coefficients than implied from their 
Concentration Indices.}
\label{ginicon}
\end{center}
\end{figure*}

Figure \ref{ginicon} shows the Gini coefficient plotted against the $E(B-V)$ 
reddening derived from the HST quasar magnitudes on the left side and the 
Concentration Index over the same $E(B-V)$ on the right side. While there is 
a weak correlation between the Gini coefficient and the reddening, we find no 
correlation between the Concentration Index and $E(B-V)$. As mentioned 
before, high Gini coefficient and low Concentration indices are an indication 
for interaction or merger, since the brightest flux pixels are highly 
concentrated in a few pixels (high G), but those pixels are not in the center 
(low C). This seems to be the case in the systems with the highest $E(B-V)$ 
in our sample, while the hosts of quasars with lower $E(B-V)$ tend to have 
both Gini coefficients and Concentration indices more consistent with those 
of normal, undisturbed galaxies \cite{abraham03}. On the lower panel of 
Figure \ref{ginicon} the Concentration Index is plotted against the Gini 
coefficient, with the unity slope shown. Only the two undisturbed galaxies 
(F2M0834$+$3506 and F2M1151$+$5359) are above this slope, with the rest 
having high Gini coefficients in relation to their Concentration Indices, 
again implying high interaction features.

While we could already see those results ``by eye'', these galaxy parameters 
are affirmation to the result that the higher the reddening in the quasar, 
the higher the chance of interaction in the quasar host. Furthermore, if the 
reddening in the spectrum fits a dust reddening template well, it is also a 
good indicator for evidence of interaction. Nonetheless, these results are 
based only on 13 highly heterogenous systems, so further observations are 
needed to improve number statistics and to confirm this claim.

\section{Notes on individual Quasars}

While most of the quasar host galaxies show some degree of irregularity and 
evidence for merger, it is worth it to look at the quasars on an individual 
basis to further inspect some peculiarities in some of the objects.

\subsection{F2M0729$+$3336}

This is the only quasar for which SDSS information was not available. Even 
though we detected only 6 photons in the X-ray observations, all of them are 
in the hard band, giving this source a hardness ratio of 1.0 and therefore 
implying a high column density.

The spectrum has strong Ca H+K absorption lines, but the spectral reddening 
does fit a dust extinction law quite well longward of about 5000 \AA\,, so 
the host galaxy contribution to the red optical light in the spectrum is not 
going to make a large impact. 

This particular system has a very compact elliptical host, with a very small 
Petrosian radius and large Gini coefficient. Beyond that however, large tidal 
tails extend out to over 20 kpc on both sides of the host, which is evidence 
for a recent merger event. This makes the total size of the host galaxy over 
35 kpc. In the $g^{'}$-band, the quasar (PSF) contribution is almost 
non-detectable, making this the system with the largest shift in 
$g^{'}$-$I_c$ color of the PSF relative to the total emission from the host. 

\subsection{F2M0825$+$4716}

This looks like a Type-2 quasar in the optical spectrum. The color of the 
quasar (PSF) would suggest that also. However, the Spex IR spectrum shows 
broad quasar lines, however and a much redder spectral slope than implied by 
the optical continuum,which is probably dominated by host galaxy emission. 
The image shows that while the central component is fit well by an elliptical 
galaxy, there is a tidal bridge linking the central galaxy to a companion 
nearby.

The optical spectrum shows double-peaked narrow emission lines, separated by 
about 600km$\, s^{-1}$. The higher redshift set has relatively strong Balmer 
lines and [O{\sc ii}]3727 emission at $z=0.803$, and probably corresponds to 
the systemic redshift of the host galaxy. The blueshifted component at 
$z=0.800$ is of higher ionization and probably corresponds to a high 
ionization outflow.

\subsection{F2M0830$+$3759}

This quasar has the lowest redshift in our sample, therefore features close 
to the nucleus will be more easily observable. The Hubble image in fact shows 
a lot of irregularity near the nucleus with shells of material along the 
major axis, either side of the nucleus. They almost cancel each other out in 
the radial profile plot, resulting in a deceptively smooth radial profile, 
while in reality the host galaxy is very disturbed.

F2M0830$+$3759 was also observed with Chandra and has a very high X-ray flux 
\citep{urrutia}. The X-ray spectrum shows a very broad Iron K$\alpha$ line, 
hinting that we are looking into the broad line region of the quasar. It has 
a moderate absorption, but a higher than usual spectral slope ($\Gamma 
\approx 2.9$). This is also the only object for which we have a gas:dust 
ratio, which is about 5 times Galactic value. While dusty, it is well within 
the 3-100 times Galactic value range deduced by \cite{maiolino}.

\subsection{F2M0834$+$3506}

The red quasar system has almost no evidence for interaction. The profile 
could fit an addition of elliptical profiles, but the morphology could also 
represent merger remnant shells. However, there are no extended features or 
tidal tails. 

It is unclear whether the blue component just to the North of the quasar 
is associated with it, or whether it is an unrelated object close to the
line of sight. There are broad H$\alpha$ and H$\beta$ emission lines 
$\sim$5000 km s$^{-1}$ blueward of the quasar's redshift, which could be 
associated with that component and/or with the quasar.

\subsection{F2M0841$+$3604}

This object shows the most irregular morphology in the host galaxy of the 
Hubble images. While the optical image clearly shows two bright nuclei, of 
which either or both could be the quasar, the radio images (from FIRST and 
also 6cm VLA A-array observations) show a steep spectrum point source right 
in between the two bright  optical components. This raises the question of 
where the active nucleus actually is.

In addition to the HST and VLA images, F2M0841$+$3604 has been observed with 
Chandra \citep{urrutia}. The Chandra photons are very hard. Unfortunately, 
the X-ray image had only 7 counts, so the position error on where the X-ray 
emission comes from is quite large. However, the Chandra photons appear to be 
spread out, so the X-ray emission could come from both nuclei. Overall, this 
is our most spectacular example of a young evolutionary merger state.

\subsection{F2M0915$+$2418}

F2M0915$+$2418 is the quasar which fits an elliptical profile best. However, 
a clear {\it tail} is seen extending to th East indication a recent 
interaction.

\subsection{F2M1012$+$2825}

This red quasar displays two nuclei only 0.15'' (1.2 kpc) apart, which would 
have been missed in ground-based observations. The two nuclei are point-like 
and especially easy to detect in the $g^{'}$-band, but the nucleus appears 
very ``smeared'' in the $I_c$-band image. The fitting of quasar and host 
galaxy is therefore impossible. The quoted luminosities for the quasar are 
only for one component, so the total nuclear luminosity is much higher than 
$M_B = -24.26$.

The rest of the host galaxy does not appear very disturbed and also here the 
central component fits an elliptical profile, were it not for that second 
nucleus. There is some blue emission north of the galaxy, but again, we do 
not know if it is associated with the quasar host.

\subsection{F2M1113$+$1244}

While the position of the quasar is quite certain, because of the clear 
stellar-like feature, it is noteworthy that the rest of the host galaxy 
consists of several bright knots, with a low surface brightness tidal tail 
extending to 30-35 kpc from the galaxy.

\subsection{F2M1118$-$0033}

The image for this red quasar shows it to be a clear merger with a nucleus 
and two compact components, one near the nucleus and the other one at the end 
of the two tidal tails trailing the system. This is the only system for which 
an exponential profile fit the host galaxy better than an elliptical (but  
only near the nucleus).

The optical spectrum for F2M1118$-$0033 shows a 4000 \AA\, Balmer break, and 
a red spectral slope which appears to be dominated by the host galaxy. The 
infrared spectrum \cite{f2m} also fails to show obvious broad emission lines, 
but does show a significant rise in K-band, probably from quasar light. 

\subsection{F2M1151$+$5359}

F2M1151$+$5359 is the quasar also fits an elliptical profile quite well, but 
there is still some residual in the host galaxy after model subtraction in 
the $I_c$-band. The blue band fits almost perfectly. So this would be 
comparable to other ``normal'', bright, blue quasars, which have an 
undisturbed elliptical galaxy as a host. The spectrum shows asymmetric 
profiles in the emission lines and narrow MgII absorption lines very close to 
the quasar redshift.

\subsection{F2M1507$+$3129}

This is the highest redshift red quasar imaged with ACS. The spectrum fits 
dust reddening quite well. Also here the central component can be well fit by 
an elliptical galaxy, but there is a low surface brightness tidal bridge 
connecting it to the red companion nearby. F2M1507$+$3129 is thus a very 
large (35 kpc) and spectacular merger at z$\sim$1. 

\subsection{F2M1532$+$2415}

HST imaging of F2M1532$+$2415 displays a truly disturbed host, with many 
components and no clear evidence of where the quasar is located among the 
constituents of the host galaxy. Radio maps show the synchrotron radiation 
coming from the central red component. They also show that this quasar is 
actually a classical Fanaroff-Riley Type II radio double (FRII, \cite{fr2}), 
with the outer components being over one arminute away from the central source.

The optical spectrum shows a very narrow H$\beta$ line and no broad MgII line, 
but in the infrared Pa$\beta$ is broad. The quasar host shows extreme 
interaction.

\subsection{F2M1656$+$3821} 

This optically-faint object shows almost only host galaxy emission in the 
optical part of the spectrum, but in the near-infrared there is a very red 
continuum and a broad Pa$\beta$ line. There are three components in the HST 
image, two point-like and one elliptical. The radio flux points to the 
emission coming from the brightest (SW) optical point source. The morphology 
of F2M1656$+$3821 could suggest that the system is a gravitational lens with 
the lensing galaxy lying between the two point sources. The reddening for 
this system might stem from the lensing galaxy and not the quasar itself, as 
is implied by the host galaxy's very red colors. However, the lack of any 
arc-like morphology in the outer optical components and the lack of spectral 
features at lower redshift in this object's spectrum argues against the 
lensing hypothesis.

\section{Conclusions}

We have observed and analyzed a sample of 13 highly reddened, luminous 
quasars from the F2M survey \cite{f2m07} with ACS in both $I_c$- and 
$g^{'}$-band to study their host galaxies. The images show interactions in 
85\% of the objects and clear evidence for mergers such as tidal tails and 
multiple nuclei. We fitted model PSF and galaxy profiles to the images. 
Within the nuclear region, most galaxies fit an elliptical profile, in 
accordance with quasar host studies such as \cite{dunlop}, but outside of the 
nuclear region this fit is no longer valid as merger features become clear.

After performing PSF-fitting, the quasar displays even redder colors and 
reddenings than implied from SDSS imaging and ESI-spectra, showing that the
optical magnitudes of the quasars are significantly contaminated by the host 
galaxies in low resolution data. Five of our galaxies show very dramatic, 
likely multiple merger morphologies, reminiscent of the simulations of 
high-$z$ quasar hosts by \cite{li}. There is a wide range in merger phase, 
with some of our quasar hosts in an apparently early part of the merger 
sequence where the merging galaxies still retain their identity 
(F2M0841$+$3604), through to later stages with multiple nuclei 
(F2M1113$+$1244; F2M1012$+$2825), and objects at the final merger stage with 
only a single nucleus and a tidal tail remaining (F2M0915$+$2418). This 
would be consistent with quasar activity being triggered relatively early in an
interaction, and continuing through the merger process. 

We calculated the Gini coefficients and the Concentration indices of the host 
galaxies and found a correlation of the Gini coefficient with reddening. The 
systems with the highest quasar $E(B-V)$ did not have high Concentration 
indices consistent with their high Gini coefficients, which implies that the 
host galaxies have bright, compact components outside of the nuclear region. 
Using this as a measure of interaction, it seems that the amount of 
interaction is weakly correlated with how obscured the quasar is.

\cite{floyd} studied a sample of 17 normal quasar hosts with HST which are 
closest to our sample in terms of luminosity and redshift. Their sample 
spanned $-24<M_V<-28$ in quasar luminosity and $0.3<z<0.42$ in redshift. 
Consistent with previous studies, only a small fraction of 
interactions/mergers were seen (they have only one object out of 17 with 
definite signatures of an interaction). However, sensitive observations of 
the hosts of unreddened quasars of similar luminosity are difficult, so it is 
possible that some signs of interaction such as faint tidal tails and 
multiple nuclei close to the quasar will have been missed. Our correlation of 
reddening with morphology argues against this, but our observed correlation 
is only weak. However, Figure \ref{nhfig} also shows obvious signs of merger 
even at nucleus to host ratios of 10. Another possible source of bias is 
that, while none of the red quasars is radio-loud, these objects are all 
radio-selected, and fall into the ``radio intermediate'' category. The fact 
that all of our quasars are disturbed in comparison to, e.g., the 
\cite{marble} sample, could be a selection effect. \cite{nano} find that 
quasars with redder colors than the SDSS composite tend to have higher 
radio-fluxes.

These caveats aside, our results may explain why only about 30\% of quasars 
with obvious signs of mergers and interactions have been found 
\citep{guyon,marble}. Reddening by dust from the host galaxy seems to be the 
most likely explanation for the redness of the F2M quasars. The host 
morphologies and colors are consistent with dusty, merging galaxies. If these 
objects were red because the line of sight just grazes the torus we would 
expect less disturbed host morphologies, in line with those of normal 
quasars. Host galaxy reddening of the quasar, whilst only mild in our cases 
(and much less than nuclear reddening by an edge-on torus) can nevertheless 
be sufficient to remove the objects from optically-selected quasar samples 
(or at best render them a small minority in such a sample). Given that these 
kind of dust-reddened Type-1 quasars like these make up about 30\% of the 
total in mid-IR selected samples, this means the fraction of quasar hosts 
associated with mergers has been significantly underestimated in the past. 
Our result is thus consistent with theories in which quasars start their 
lives obscured by dust, and only appear in optical or soft X-ray 
surveys after the dust along the line of sight to the nucleus has been 
cleared by quasar winds \citep{sanders,sanders96,springel,hopkins07}.

\acknowledgments

We thank Elinor Gates for assistance with writing the {\em fithost}
program. We also thank Wim de Vries for help with the ACS astrometry.

This work was partly performed under the auspices of the US Department of 
Energy, National Nuclear Security Administration by the University of 
California, Lawrence Livermore National Laboratory under contract No. 
W-7405-Eng-48. Funding for this project was supplied by Hubble Grant
10412.

This publication makes use of data products from the Two Micron All Sky 
Survey, which is a joint project of the University of Massachusetts and the 
Infrared Processing and Analysis Center/California Institute of Technology, 
funded by the National Aeronautics and Space Administration and the National 
Science Foundation.

This publication makes use of the Sloan Digital Sky Survey. Funding for the 
SDSS and SDSS-II has been provided by the Alfred P. Sloan Foundation, the 
Participating Institutions, the National Science Foundation, the U.S. 
Department of Energy, the National Aeronautics and Space Administration, the 
Japanese Monbukagakusho, and the Max Planck Society, and the Higher Education 
Funding Council for England. The SDSS Web site is http://www.sdss.org/.


\end{document}